\documentclass[12pt]{article}
\usepackage{epsfig, amssymb, graphicx, amsmath} 
\usepackage{amsthm}
\usepackage[round-mode=places, round-integer-to-decimal, round-precision=2,
 table-format = 1.2, 
 table-number-alignment=center,
 round-integer-to-decimal,
 output-decimal-marker={,}
 ]{siunitx} 
\usepackage{cleveref}
\usepackage{booktabs}
\usepackage{enumitem}
\usepackage{eurosym}	
\usepackage{wrapfig}	
\usepackage{float}		
\usepackage{footnote}
\usepackage{subcaption}
\usepackage{amsthm}
\usepackage{thmtools}
\usepackage{bbm}		
\newcommand{\ImageWidth}{16cm}
\usepackage{tikz}
\usetikzlibrary{decorations.pathreplacing,positioning, arrows.meta}
\usepackage[authoryear]{natbib}
\bibpunct{(}{)}{,}{a}{}{,}

\makeatletter
\def\ps@pprintTitle{%
	\let\@oddhead\@empty
	\let\@evenhead\@empty
	\def\@oddfoot{}%
	\let\@evenfoot\@oddfoot}
\makeatother

\newcommand{\be}{\begin{equation}}
\newcommand{\en}{\end{equation}}

\begin{document}
	
\begin{savenotes}
\title{
\bf{ 
Neural Network \\
Middle-Term Probabilistic Forecasting
of \\ Daily Power Consumption}}
\author{
Michele Azzone$^\ddagger$ \& 
Roberto Baviera$^\ddagger$ 
}

\maketitle

\vspace*{0.11truein}
\begin{tabular}{ll}
$(\ddagger)$ & Politecnico di Milano, Department of Mathematics, 32 p.zza L. da Vinci, Milano \\
\end{tabular}
\end{savenotes}

\vspace*{0.11truein}
\begin{abstract}
\noindent
 	Middle-term horizon (months to a year) power consumption prediction is a main challenge in the energy sector, 
 	in particular when probabilistic forecasting is considered.
	We propose a new modelling approach that incorporates trend, seasonality and 
	weather conditions, as explicative variables in a shallow Neural Network with an autoregressive feature.
	
	\noindent
 	We obtain excellent results for density forecast on the one-year test set applying it to the daily power consumption in New England U.S.A..
	The quality of the achieved power consumption probabilistic forecasting has been verified, 
	on the one hand, 
	comparing the results to other standard models for density forecasting
	and, on the other hand, 
	considering measures that are frequently used in the energy sector as pinball loss and CI backtesting. 
\end{abstract}

\vspace*{0.11truein}
\vspace*{0.11truein}

\vspace*{1cm}
\begin{flushleft}
Cite as: Baviera, R. \&  M. Azzone (2021). Neural network middle-term probabilistic forecasting of daily power consumption. \textit{Journal of Energy Markets}, \textit{14.1}, 1-26. \\
\end{flushleft}

\vspace{2cm}
\begin{flushleft}
{\bf Address for correspondence:}\\
Roberto Baviera\\
Department of Mathematics \\
Politecnico di Milano\\
32 p.zza Leonardo da Vinci \\ 
I-20133 Milano, Italy \\
Tel. +39-02-2399 4575\\
Fax. +39-02-2399 4621\\
roberto.baviera@polimi.it
\end{flushleft}

\newpage

\begin{center}
\Large\bfseries 
Neural Network Middle-Term Probabilistic Forecasting\\
of Daily Power Consumption
\end{center}

\vspace*{0.21truein}

\section{Introduction}
Power consumption forecasting is crucial in the energy sector. 
In particular, 
middle-term forecasting, i.e. in a time-horizon between a few months and a year\footnote{
As standard in the literature, we refer to short-term consumption forecasting as the 
prediction of the consumption 
over an interval ranging from minutes up to few weeks and to long-term consumption forecasting as the prediction for time-horizons over one year.}, 
plays a key role in the planning of power systems both for future generation plants and for transmission grids \citep[see, e.g.,][]{HongFan2016}.
Furthermore, a good power consumption forecasting has also consequences for
network reliability and leads to a reduction in energy costs and carbon emissions.

Utilities and grid operators prefer to obtain a probabilistic forecasting of the power consumption rather than a point forecasting.
A probabilistic forecasting does not provide only the future expected value but also informs on the expected distribution.
This technique gained momentum in the energy sector after the Global Energy Forecasting competition 2014
(GEFCom) on a power data-set of New England in the U.S.A. 
\citep[see, e.g.,][and references therein]{HongFan2016, NowotarskiWeron2018}. 
This competition has been followed by another one in $2017$; 
both have attracted participants from industry and academia, awarding the best hierarchical forecasting and probabilistic forecasting model.
In this study we focus on the GEFCom 2017 data-set. 

The key issue in model selection is 
to identify the correct features of the problem and their relation with power consumption; 
this allows understanding and hedging the risks that arise from unexpected jumps in the consumption.
We are interested in the impact of weather conditions on power consumption; 
they play the most relevant role in middle-term forecasts compared to economic 
and demographic drivers that play a role in longer forecasts \citep[see, e.g.,][]{HyndmanFan2010}. 
We desire to model the dependency from weather conditions over middle-term;
the most natural technique, now standard in power consumption forecasting, is known as \textit{ex-post} forecasting. 
It has been applied to middle-term probabilistic forecasting of power consumption on the French distribution network \citep{GNK2014} and 
on the National Electricity Market of Australia \citep{HyndmanFan2010}.

We use a machine learning technique. 
These techniques have been shown to provide interesting results 
in short-term point consumption forecast
\citep[see, e.g.,][]{FanChen2006, Meng2009, Shi2018}. 
In this paper we consider, as machine learning technique, a Neural Network (hereinafter NN).
Most commonly, for each time step, a NN returns the expected value, i.e. a single point forecasting and it is therefore referred to as a point-forecasting method.
As already mentioned, market players consider more interesting 
probabilistic forecasting methods that provide information on the distribution of future values. 
However, the literature on probabilistic forecasting of electrical consumption via NN is quite recent and rather limited.
This technique has been considered only by \citet{VFM2018} for one-hour density forecasting at household level.
The idea is to have, as NN output, the vector of parameters of the density instead of the single point forecasting 
and to maximise the logarithm of the likelihood. 
This idea was introduced in financial time-series by \citet{OrmoneitNeuneier1996}, 
but since then it has seldom been considered in the financial and energy literature \citep[see, e.g.,][]{FKG2010wind, NTS2013}. 
We consider the simple case of a Gaussian density forecasting following the results achieved in \citet{BavieraMessuti2019} using Gaussian Processes 
for density forecasting of power consumption of one operator in North-East England. These results for Gaussian density forecasts 
look promising both in terms of sharpness and reliability.

\bigskip

In this study, we apply a NN to a density forecasting for middle-term consumption up to one year:
this is the first difference w.r.t. \citet{VFM2018}; they focus on short-term forecasting.
Moreover, the main successes of NN have been obtained on big data-set, where this technique is reliable and efficient \citep[see e.g.][and references therein]{lecun2015}.
The real challenge is to use these techniques with small data-sets. 
This is a common demand in this industrial sector due to the rapid changes that are observed in the electricity market. 
To show the effectiveness and the quality of the proposed technique,
we consider the extreme case where we train our model on a few years' 
 data-set and predict daily power consumption on a one-year horizon. 
To the best of our knowledge, the use of NN for middle-term density forecasting is new in the literature.

In this way, we can compute the densities of the consumption forecast for the proposed hybrid model obtaining excellent performances
compared to standard benchmark models 
\citep[as the ARX, see e.g,][p.534 et seq.]{box15}
and to other machine learning approaches \citep[as the Gaussian Process, see e.g.][]{Rasmussen06}. 
The evaluation of models performances is obtained by comparing the forecasted results and the realised consumption over one-year test set. 
To value the quality of the forecasting, we consider techniques that are standard either in the energy sector, as pinball loss \citep[see, e.g.,][]{NowotarskiWeron2018}, or in the banking sector after the introduction of Basel II Accords, 
as backtesting \citep{Kupiec1995, Christoffersen1998}.

\bigskip

The main contributions of the paper are threefold.
First, we introduce a hybrid model that joints the advantages of
classical univariate time-series analysis and a shallow NN. 
In particular, via a NN we incorporate 
in power consumption density forecasting the dependency from weather conditions and from previous times: 
the network structure we propose is new in the literature and it is designed to obtain both density forecasting and the autoregressive feature observed in the time-series. 
Second, we show that a NN technique relying on a small data-set -- with only a few year-long training window-- achieves an excellent
forecasting of power consumption using only weather data in middle-term forecasts.
Third, we value the density forecasting via some sharpness and reliability measures, 
showing the quality of the achieved results. In particular, we show that results are not only accurate
but also robust over-time with a method designed for machine learning techniques.

\bigskip

The rest of the paper is organised as follows. 
In Section \ref{sec:dataset}, we summarise the key characteristics of the data-set we analyse. 
In Section \ref{sec:methodology}, we present the methodology; in particular, we describe in detail i) the proposed model 
and how the weather conditions are introduced in the NN, ii) the forecasting technique and the evaluation methods. 
Sections \ref{sec:results} and \ref{sec:exante} show the main numerical results and Section \ref{sec:conclusions} concludes.

\section{Data-set Description}
\label{sec:dataset}

New England is a region composed of six states in the Northeastern U.S.A.. 
In 2010 New England population accounted approximately for the 5\% of U.S.A. population. 
We use the GEFCom2017 data-set on New England households' consumption. 
It contains the data published by the overseer of New England bulk electric power system, ISO New England (ISONE). 
The ISONE load data (on households' consumption) includes 
the aggregated consumption of the whole New England area and
eight different New England zones: every zone corresponds to a state, except the state of Massachusetts, composed of three zones. 
The data-set is available via the ISONE zonal information page and is updated periodically; at the time of the GEFCom2017 the data-set 
included data 
up to March 2017. 
More than 300 academy and industry professionals, divided into 177 teams, participate in the 2017 competition. 
The competition features two different tracks: the open data track and the defined data track. We use the data-set of the defined data track.
The load data contains aggregated households' hourly consumption in MWh together with averaged wet bulb and dry bulb hourly temperatures 
in Fahrenheit degrees. 
We work with daily data for the whole New England area. 
We aggregate the power consumption values and average the weather conditions, i.e. we consider the daily consumption and the average temperature for every day.
The weather data-set represents the daily average of hourly records of weather conditions for the whole New England region. It includes two different variables:
	\begin{itemize}[noitemsep, nolistsep]
		\item dry bulb temperature, in $^{o}F$;
 \item wet bulb temperature, in $^{o}F$.
	\end{itemize}

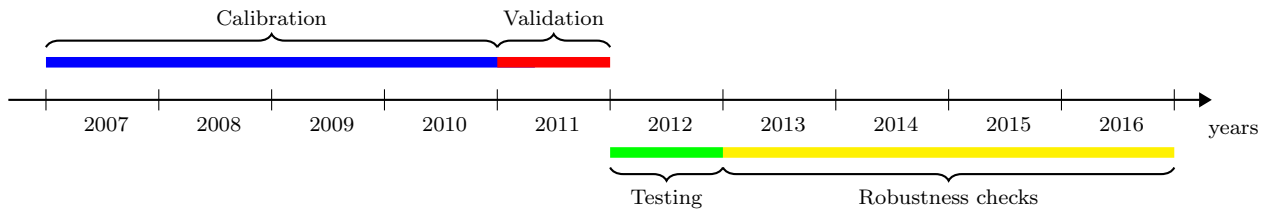
\begin{figure}[h!]
	\begin{center}
    \begin{tikzpicture}
\draw[thick, -Triangle] (0,0) -- (\ImageWidth,0) node[font=\scriptsize,below left=5pt and -22pt]{years};

\foreach \x in {0.5,2,3.5,5,6.5,8,9.5,11,12.5,14,15.5}
\draw (\x cm,4pt) -- (\x cm,-4pt);

\foreach \x/\descr in {1.3/2007,2.8/2008, 4.3/2009, 5.8/2010,7.3/2011,8.8/2012,10.3/2013,11.8/2014,13.3/2015, 14.8/2016}
\node[font=\scriptsize, text height=1.75ex,
text depth=.5ex] at (\x,-.3) {$\descr$};

\draw[blue, line width=4pt] (0.5,.5)  --  +(6.5,0);
\draw[line width=4pt, red] (6.5,.5) --  +(1.5,0);

\draw[line width=4pt, green] (8,-.7) --  +(1.5,0);
\draw[line width=4pt, yellow] (9.5,-.7) --  +(6.0,0);

\draw [thick ,decorate,decoration={brace,amplitude=5pt}] (0.5,0.7)  -- +(6.0,0) 
       node [black,midway,above=4pt, font=\scriptsize] {Calibration};
\draw [thick ,decorate,decoration={brace,amplitude=5pt}] (6.5,0.7)  -- +(1.5,0) 
       node [black,midway,above=4pt, font=\scriptsize] {Validation };
\draw [thick,decorate,decoration={brace,amplitude=5pt}] (9.5,-.9) -- +(-1.5,0)
       node [black,midway,font=\scriptsize, below=4pt] {Testing };
\draw [thick,decorate,decoration={brace,amplitude=5pt}] (15.5,-.9) -- +(-6.0,0)
       node [black,midway,font=\scriptsize, below=4pt] {Robustness checks};
\end{tikzpicture}
\end{center}
\caption{\small The timeline describes the data segmentation that we use to calibrate, validate, test and check the robustness of the proposed model. 
We calibrate the model on the years before 2011 and select the best model hyper-parameters on the validation set (2011). 
We test the performance of the model on 2012 and check its robustness on the years 2013-2016.} 
\label{fig:Sketch_data}
\end{figure}

We select $10$ calendar years of the GEFCom2017 data-set, from January $2007$ up to December $2016$.
We use the year 2011 to validate the calibrated model and select the best hyper-parameters,
the year $2012$ for testing and the years from $2013$ to $2016$ 
to verify the robustness and reliability of the results achieved with the selected model. Years from $2007$ to $2010$ are used for calibration. 
The timeline in Figure \ref{fig:Sketch_data} describes the data segmentation. 

Table \ref{table:DescrStatsWeather} contains descriptive statistics about daily power consumption and weather data on the 2007-2010 time window. 
	\begin{table} [!h] 
		\centering 
		{\footnotesize \begin{tabular}{|c|c|c|c|c|c|} 
			\hline 
			& Min & Max & Mean & Median & Standard Deviation \\ 
			\hline \hline 
			Consumption [$GWh$] &270.29 & 520.28 &335.08 & 346.20 & 41.09 \\ \hline \hline 
			Dry bulb [$^{o}F$] & 4.80& 85.83 &49.66 &50.94 &17.73 \\ 
			\hline 
			Wet bulb [$^{o}F$] & -13.04 & 72.46& 37.80 & 38.56 & 18.91 \\ 
			\hline 
	 
		\end{tabular} } 
		\caption{\small{Descriptive statistics for cumulative daily power consumption and average daily weather data in New England for the 2007-2010 time window.}} 
		\label{table:DescrStatsWeather} 
	\end{table} 
We remove the $29^{th}$ of February to preserve the seasonality structure in leap years.

\begin{figure}[h!]
			\centering
		\hspace{-1.2cm}
		\includegraphics[width=0.9\textwidth]{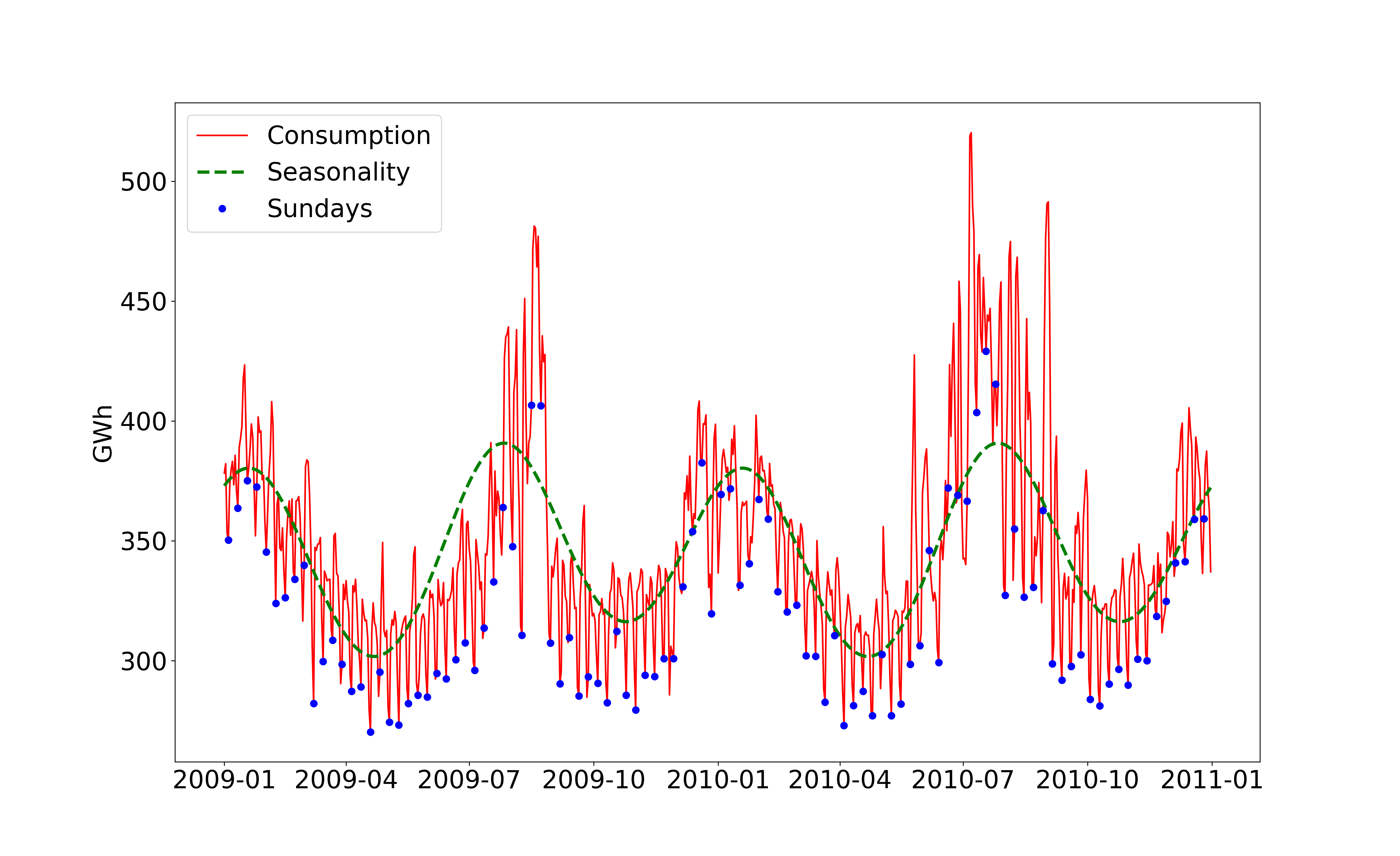}
		\caption{\small{Cumulated power consumption in New England (red line) between January 2009 and December 2010 with the fitted seasonal behaviour (green dashed line). One can notice a yearly seasonal behaviour (with two peaks per year in summer and winter) and locally lower values on Sundays 
(marked with the blue circles).}}
		\label{fig:powerconsumption}
\end{figure}

The yearly and weekly seasonality of power consumption is noticeable in the data-set.
In Figure \ref{fig:powerconsumption} we plot two years (2009 and 2010) of electricity consumption. 
We infer a yearly seasonal behaviour with two peaks per year (winter and summer). 
The peak in summer is due to air-conditioning while the one in winter to heating.

We observe that 
i) power consumption in summer is double w.r.t. power consumption during spring or autumn, 
ii) consumption on weekdays is higher than 
 on Sundays and
iii) volatility (and spikes) is larger during summer than in the rest of the year.
These are well-known {\it stylised facts} common to most households' electricity consumption time-series,
specifically in regions where there is massive use of air-conditioning during summer and a significant source of heat comes from heat-pumps during winter. 
For these reasons, and, in particular, due to the seasonality in the observed volatility, it is now standard in the literature to model the logarithm of 
households' electricity consumption \citep[see, e.g.,][for a review]{HongFan2016}. 

 In the next section we describe in detail the adopted methodology.
\section{Methodology}
\label{sec:methodology}

\subsection{The model}

The great majority of the forecasting studies on the power market is focused on price forecasting \citep[see, e.g.,][for a review]{NowotarskiWeron2018}; 
in this paper we focus on a middle-term density forecast for power consumption, where the number of studies in the literature is rather limited \citep[see, e.g.,][]{HongFan2016}.
We aim to forecast 
future power consumption over the middle-term horizon, modelling both seasonal and weather-related features;
the goal is to obtain a density forecasting of daily consumption
with a reasonably simple description.

\bigskip

As already mentioned in the previous section, 
we model log-scaled daily electricity consumption data, as it is standard in the literature for the probabilistic forecasting in the energy sector.
The characteristics of power demand we desire to model are,
on the one hand, long-term trend and seasonal behaviour (both yearly and weekly), and on the other hand, daily autocorrelation, the relation with weather conditions and
variance clustering.
We build a hybrid model composed of two blocks:
a linear model that describes trend and seasonality and
a NN that describes the weather influence and 
the nonlinear effects of seasonality (e.g. the calendar effect on the variance).

A hybrid model approach where one separates first
trend and seasonality, and then analyses separately the residuals 
 is standard in the energy literature
\citep[see, e.g.,][and references therein]{Benth2008}. 
First, the relation between the logarithm of consumption and calendar variables is established through a General Linear Model (GLM).
Then, we investigate the relation between GLM residuals and weather variables through a shallow 
NN. 

\bigskip

The GLM part of the model is elementary.
We model the natural logarithm of power consumption $Y_t$ as 
\begin{equation}
 	Y_t = T_t + S_t + R_t \quad ,
\label{eq:model} 
\end{equation}
where
\begin{equation*}
		\begin{cases}
		T_t &= \beta_0 + \beta_1 t \\
		S_t &=\sum_{k=1}^{2}\left[\beta_{1+k} \sin\left(k\omega t\right) +\beta_{2+k}\cos\left(k\omega t\right)\right]+ \beta_6D_{Sat}(t) + \beta_7 D_{Sun}(t) +\beta_8 D_{Hol}(t)
		\end{cases} \quad ,
\end{equation*}
and calendar  time is measured via the cardinality $t$ of the observation,
starting from 0 on the first date in the data-set;
$ R_t $ are the residuals,
$T_t$ the trend term and $S_t$ the seasonality both yearly and weekly.
For the yearly seasonality we use both annual and semiannual seasonality terms with $\omega := 2 \pi/365 $;
these sinusoidal terms with the order $k$ up to two describe the two yearly peaks observed in Figure \ref{fig:powerconsumption}.
Moreover, the holidays' contribution is modelled with a dummy variable.
The weekly seasonality is introduced via two dummy variables for Saturday and Sunday. 

\bigskip

\begin{figure}[h!]
	\centering
	\hspace{-1.2cm}
	\includegraphics[width=0.5\linewidth]{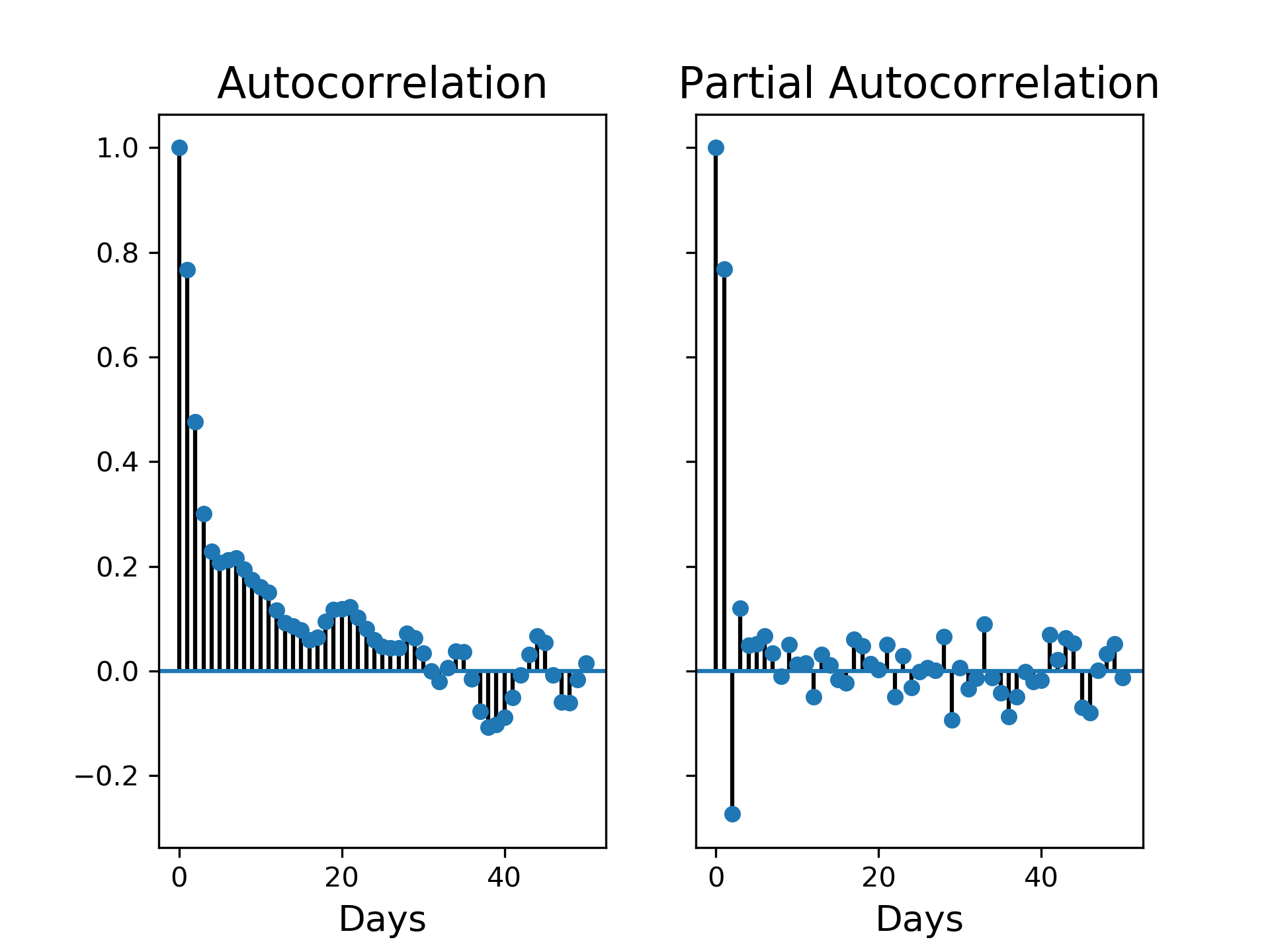}
	\caption[Autocorrelation function and partial autocorrelation function 
of log-additive deseasonalized consumption]{\small Autocorrelation function and partial autocorrelation function 
of seasonally adjusted consumption time-series in the time-window 2007-2010. 
We observe that the time-series is autocorrelated; 
from the partial autocorrelation we notice that the time-series is at least AR(1). }
\label{fig:acf_pacf}
\end{figure}

We measure the autocorrelation and the partial autocorrelation of the residuals $R_t$ on the time window 2007-2010; 
Figure \ref{fig:acf_pacf} highlights the necessity of a one day auto-regressive component.
Furthermore, the augmented Dickey-Fuller test refuses the null hypothesis of a unit root for the deseasonalized time-series.

Residuals $R_t$ are modeled via a NN.
NN are generally used for point forecasting where they can achieve excellent results.
However, the literature on probabilistic forecasting of electrical consumption with NN is scarce. 
As already mentioned in the introduction, there is only 
one study yet available \citep{VFM2018}: 
the authors use a standard Feed-Forward NN (FFNN) for density forecasting. 

A FFNN represents the most standard and straight-forward type of NN. 
Its name arises from the fact that, in this structure, the information moves only in one direction, forward, from the input layer through the hidden neurons and finally to the output layer. Furthermore, if we represent 
a FFNN architecture with a computational graph, 
where each node represents a neuron and each edge represents the information flow between two neurons, 
we would see that those connections do not form any cycle.

However, because the analysed time-series show a significant one-time-step autocorrelation, 
we consider a NN with an “autoregressive” feature: 
NN outputs at time $t-1$ are also used as two additional inputs to the NN for the next time in the series. 
This feature resembles the structure of Recurrent NN where the output of a given layer is considered an input to the same NN layer 
\citep[see, e.g.,][]{RCF2014, Kong2017};
in this paper we consider only a feedback from the output layer into the input layer. 
This rather simple architecture is shown in Figure \ref{fig:nn_scheme}.
This construction replicates the “autoregressive” feature in standard AR models. 
Because the structure resembles the one of AutoRegressive eXogenous models (ARX), 
we call the proposed hybrid model a NN with Autoregression and eXogenous inputs (NAX).
In the next subsection we describe in detail the NN part of the model.

\subsection{Neural Network with Autoregression and eXternal inputs}
\label{subsec: NAX}

It is well-known that, after having detrended and deseasonalized the time-series, 
the impact on power consumption of weather conditions in general and of temperature in particular cannot be neglected \citep[see, e.g.,][]{HongFan2016}.

\begin{figure*}
			\centering
		\hspace{-1.2cm}
		\includegraphics[width=0.8\textwidth]{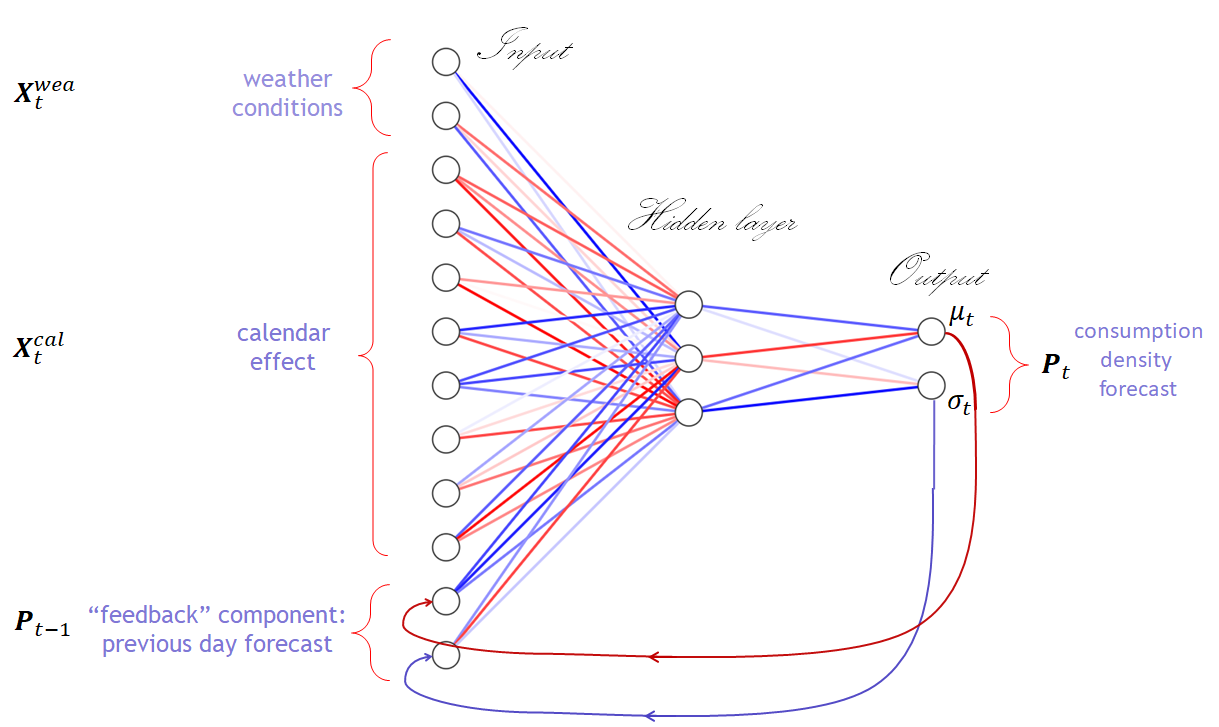}
		\caption[]{\small 
Scheme of the NAX model selected in the validation procedure. Notice a twelve-dimensional input 
(two weather conditions $\boldsymbol{X}^{wea}_t $, eight neurons for the calendar effects $\boldsymbol{X}^{cal}_t$  and 
two auto-regressive terms $\boldsymbol{P}_{t-1} $), 
a three neuron hidden layer and a two-dimensional output $\boldsymbol{P}_t $ with two components ($\mu_t$ and $\sigma_t$). }
\label{fig:nn_scheme}
\end{figure*}

In this paper, residuals of power consumption
are modelled via 
 a simple NN with 
a feedback from the output to the input.
Moreover, following the promising results obtained by Gaussian Processes in probabilistic forecasting of power consumption \citep[see, e.g.,][]{BavieraMessuti2019}, 
we consider the residuals $R_t$ normally distributed with unknown mean $\mu_t$ and variance $\sigma_t^2$.
Considering a NN with these three characteristics is the main contribution of this study from a modelling perspective:
i) the NN  
presents a two-dimensional output vector that models time-dependent density parameters,
ii) it presents an ``autoregressive" structure with NN output included as input at the next time step and
iii) it incorporates the information coming from weather conditions and from calendar effects.
Figure \ref{fig:nn_scheme} describes an example of the NAX structure; 
we observe a vector output $\boldsymbol{P}_{t}$ with two components ($\mu_{t}$ and $\sigma_{t}$) and 
inputs at time $t$ that include both the exogenous inputs ($ \boldsymbol{X}_t $) at time $t$ and the outputs ($\boldsymbol{P}_{t-1}$) at time $t-1$.

The NN we consider has $I$ exogenous inputs (including both weather conditions and calendar effects) and one hidden layer with $N$ 
neurons.\footnote{Notice that, in this case, the NN reduces to a Recurrent NN with one nonlinear hidden layer and one linear output layer.}
Thanks to the simplicity of the structure it is possible to describe this NN with the equation 
\begin{equation}
 \boldsymbol{P}_t = 
\boldsymbol{l} \, {\cal H}\left(\boldsymbol{w} \, \boldsymbol{X}_t+\boldsymbol{f} \, \boldsymbol{P}_{t-1}+\boldsymbol{w}_0 \right) + \boldsymbol{l}_0 \;\;, 
\label{equation:nn}
\end{equation}
where 
\begin{itemize}
\item ${\cal H}$ is the nonlinear activation function (e.g. softmax, sigmoid), a vector in $\mathbb{R}^N$;
\item $\boldsymbol{l}$ is a $2 \times N$ array of weights for the output of the $N$ neurons in the hidden layer, while $\boldsymbol{l}_0$ is the two-dimensional bias on the output;
\item $\boldsymbol{w}$ is the $N \times I$ array of weights for the input of the $N$ neurons in the hidden layer, while $\boldsymbol{w}_0$ is the $N$-dimensional vector of bias on the $N$ neurons;
\item $\boldsymbol{f}$ is the $N \times 2$ array of weights of the feedback term of the $N$ neurons.
\end{itemize}

The analysis is divided into four main steps, the first three in line with NN literature \citep[see, e.g.,][and references therein]{Ripley2007pattern}, and 
a fourth one to evaluate the robustness of the proposed solution.
We implement the NAX in Keras \citep[see, e.g.,][]{chollet2015keras} with TensorFlow backend. 
To avoid the problem of different input and output scales, 
we normalize every input and output data with the min-max normalization technique 
\citep[see e.g.][p.1]{patro2015normalization} before feeding them to the NAX. 
Every column of the data matrix is standardised and de-standardised using only information in the training set.

First, we train the NAX
maximizing the Gaussian likelihood function as proposed by \citet[][eq. 2]{VFM2018}, where
\[
L(\mu_t,\sigma_t | R_t)= \frac{1}{\sqrt{2\pi \sigma_t^2}}\exp{-\frac{\left(R_t-\mu_t\right)^2}{2\sigma_t^2}}\;\;.
\] 
is the likelihood function for the residual $R_t$ in equation (\ref{eq:model}).\footnote{Notice 
that the NAX outputs are not a direct estimation of the residual $R_t$ but 
of the two parameters $\mu_t$ and $\sigma_t$. 
When training the NN we look for the parameters that maximize the likelihood of the realised residuals in the training set.} 

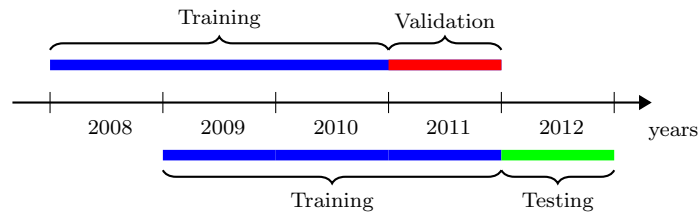
\begin{figure}[h!]
	\begin{center}
    \begin{tikzpicture}
\draw[thick, -Triangle] (1,0) -- (9.5,0) node[font=\scriptsize,below left=5pt and -22pt]{years};


\foreach \x in {1.5,3,4.5,6,7.5,9}
\draw (\x cm,4pt) -- (\x cm,-4pt);


\foreach \x/\descr in {2.3/2008, 3.8/2009, 5.3/2010,6.8/2011,8.3/2012}
\node[font=\scriptsize, text height=1.75ex,
text depth=.5ex] at (\x,-.3) {$\descr$};

\draw[blue, line width=4pt] (1.5,.5)  --  +(6,0);
\draw[line width=4pt, red] (6,.5) --  +(1.5,0);

\foreach \x/\perccol in
{3/100,4.5/75,6/0}
\draw[blue, line width=4pt] 
(\x,-.7) -- +(1.5,0);
\draw[line width=4pt, green] (7.5,-.7) --  +(1.5,0);

\draw [thick ,decorate,decoration={brace,amplitude=5pt}] (1.5,0.7)  -- +(4.5,0) 
       node [black,midway,above=4pt, font=\scriptsize] {Training };
       \draw [thick ,decorate,decoration={brace,amplitude=5pt}] (6,0.7)  -- +(1.5,0) 
       node [black,midway,above=4pt, font=\scriptsize] {Validation };
\draw [thick,decorate,decoration={brace,amplitude=5pt}] (9,-.9) -- +(-1.5,0)
       node [black,midway,font=\scriptsize, below=4pt] {Testing };
\draw [thick,decorate,decoration={brace,amplitude=5pt}] (7.5,-.9) -- +(-4.5,0)
       node [black,midway,font=\scriptsize, below=4pt] {Training };

\end{tikzpicture}

\end{center}
	\caption{\small An example of the timeline that describes the data segmentation that we use to validate 
(above the time axis) and test (below the time axis) the NAX model. The length in years of the training window is an hyper-parameter;
we train on the years before 2011 and select the best model hyper-parameters on the validation set (2011): 
in particular, we show in Figure the case of a 3 year training set (2008-2010). 
Then, for testing we calibrate the selected model on the years before 2012 (keeping the training set length equal to the selected hyper-parameter) 
and we test the performance of the model on 2012. } 
\label{fig:Sketch_data_2}
\end{figure}

Second, we select the hyper-parameters of the NAX model. 
We perform a grid search on the hyper-parameters on the validation set (the year 2011). 
See Figure \ref{fig:Sketch_data_2} for a description of the data segmentation in training, validation and testing set.
For every possible combination of the hyper-parameters in Table \ref{table:Hyperparam},
NAX model is trained on the training set and
the model with the best Root Mean Squared Error (RMSE) on the validation set is selected. 
Minimising RMSE is standard among power suppliers: this choice helps to focus calibration on periods of higher consumption, 
often characterised by higher electricity costs. 
Even if we estimate the network w.r.t. the Gaussian likelihood (in order to have an estimation of the consumption density)
 we decide to select the model that satisfies this industry requirement.
In this study, the training set time-window length is one of the hyper-parameters. 
As already discussed in the introduction, due to the rapid changes that are observed in the 
power market, it is important to avoid training on years that do not help the NN in its ability to generalise: 
it has been selected as the longest time-window for which one observes a decreasing RMSE in the validation set. 
In Figure \ref{fig:Sketch_data_2} we show an example of the data segmentation that we use to validate the NAX model (cf. training and validation sets above the time axis):
in particular, in the Figure we consider the case when the training set time-window is 3 years.

\begin{table} [!h] 
		\centering 
		{\footnotesize \begin{tabular}{|l|r|} 
			\hline 
			{\it Hyper-parameter} & {\it Values} \\ 
			\hline
			Number of neurons (hidden layer) &3, 4, 5, 6, 8, 10\\
			\hline 
			Activation function & softmax and sigmoid \\
			\hline
			Initial learning rate &0.1, 0.01, 0.001, 0.0007, 0.0005, 0.0001\\
			\hline
			Batch size & 50, 100, 350, no batch\\
			\hline
			Regularization parameter & 0.01, 0.001, 0.0001, 0\\
			\hline
	 	 Training set time-window & 1, 2, 3 and 4 years\\
			\hline 
		\end{tabular} } 
		\caption{\small Set of hyper-parameters analysed in the validation step.} 
		\label{table:Hyperparam} 
	\end{table} 

 
Third, we consider the test set (the year $2012$), on which we perform the model evaluation 
according to a set of indicators. 
In Figure \ref{fig:Sketch_data_2} we also show an example of the data segmentation that we use to test the NAX model (cf. training and testing sets below the time axis):
we train the selected model on the three years before 2012 (training set length equal to the selected hyper-parameter) 
and test the performance of the model on 2012.
Both validation and testing are conducted via an
{\it ex-post} probabilistic forecasting of power consumption, standard in middle-term forecasting.
The next subsection describes in detail the {\it ex-post} density forecasting technique and the evaluation methods we consider in probabilistic forecasting.

Finally, in addition to the three classical steps, we perform a robustness test to verify the quality of the proposed hybrid model: 
we evaluate model performances in the following years $2013$-$2016$ (cf. also Figure \ref{fig:Sketch_data}), 
training the model with the same hyper-parameters selected in the validation set. 
A summary of the results is discussed in Section \ref{sec:results}.

\subsection{The ex-post forecasting technique and the evaluation methods}

\begin{figure}[h!]
		\centering
		\includegraphics[width=.7\linewidth]{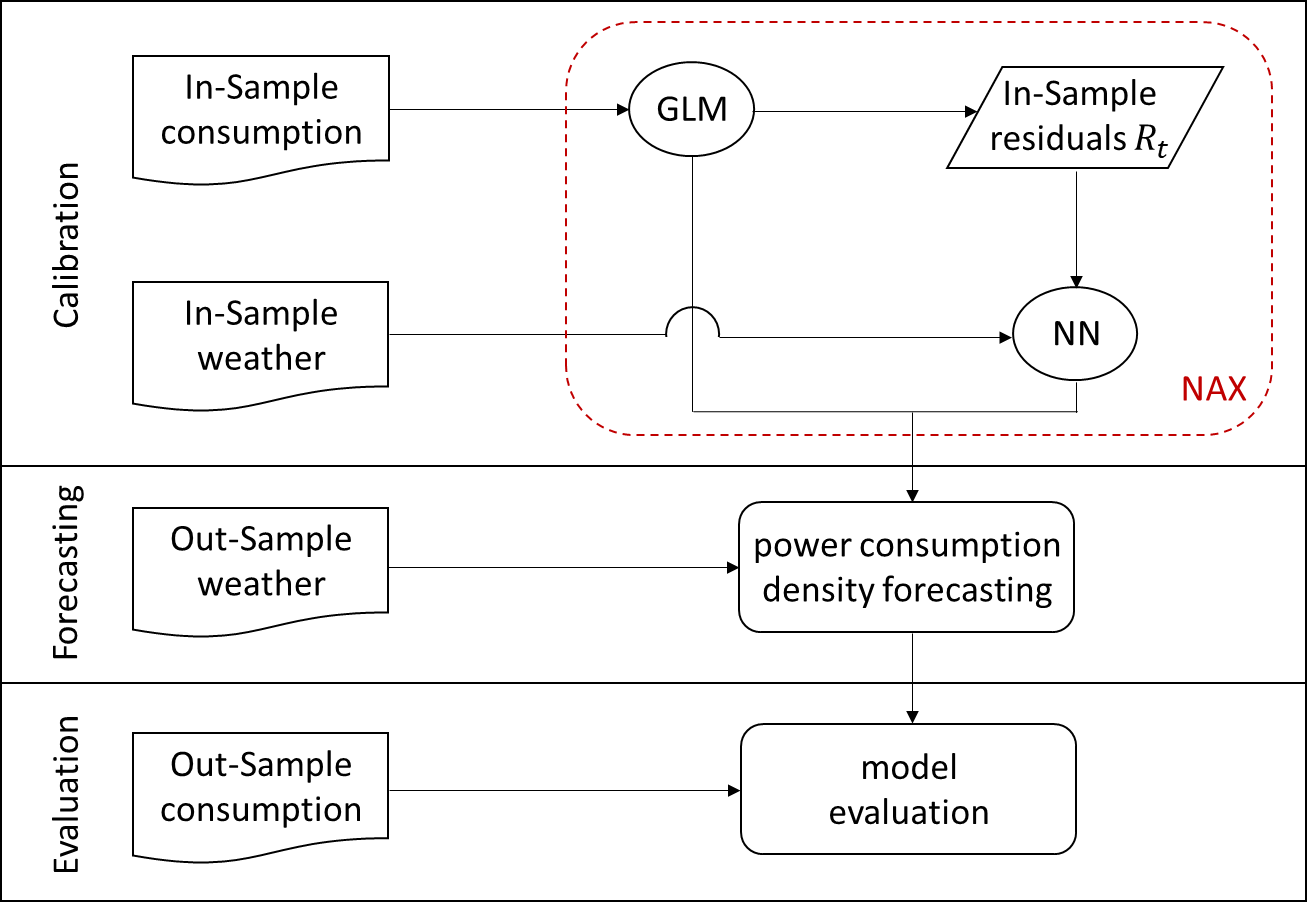}
		\caption[Flow-diagram]{\small Flow-diagram of the three stages of the {\it ex-post} technique: calibration, forecasting and evaluation of the proposed model. 
The presence of a forecasting stage is the main difference w.r.t. the standard technique.}
\label{fig: flowchart}
\end{figure}

The probabilistic forecasting of middle-term daily power consumption 
is obtained via {\it ex-post} forecasting. This technique, introduced by \cite{HyndmanFan2010} in the power consumption sector, 
is commonly used in middle to long term power consumption forecasts \citep[see, e.g.][]{GNK2014}.
The method is divided into three stages \citep[see, e.g.,][p.1144]{HyndmanFan2010} shown 
in the flow diagram of Figure \ref{fig: flowchart} for the model we consider: calibration, forecasting and evaluation.
The first stage is applied to the In-Sample (training) set. The second and the third stages use the Out-of-Sample set:
it can be either the validation set or the test set, depending on whether we are selecting the optimal hyper-parameters or we are testing the model. 

In the calibration stage, the NAX model --in its two components GLM and NN-- is trained on the In-Sample set, with both power and meteorological data.
GLM is calibrated through Ordinary Least Square and the NN is trained using the Keras ADAM optimizer.

In the forecasting stage, the density is estimated via an {\it ex-post} forecast. This second stage is the main difference with standard forecasting.
This forecasting uses the weather conditions
in the Out-of-Sample set 
 to forecast the power consumption; as well explained by
\citet[][p.443]{GNK2014} it
``allows
us to quantify the performances of our model without embedding the meteorological
forecasting errors''.
The idea is that,
to focus on the ability of the model to describe a strong and reliable relation between temperatures and power consumption,
one supposes to know perfectly the temperatures in the Out-of-Sample period.
With the proposed model an {\it ex-post} forecasting is straightforward. 

Finally, in the evaluation stage, 
the quality of the model forecasting is evaluated comparing it
with the Out-of-Sample realised test consumption data. 
The model is evaluated both 
in terms of point consumption forecasting and 
in terms of sharpness and reliability of the predicted densities.

\bigskip

Besides the standard measures of point consumption forecasts as RMSE and Mean Absolute Percentage Error (MAPE), 
we provide some evaluation methods of density forecasting.
When dealing with density forecasting
it is more difficult to value the quality of the forecasting. We are not able to observe the realised distribution of the underlying process: 
we cannot compare the predicted distribution to the true one, as we only have one realisation for each distribution.
The evaluation is based on two main measures: 
the \textit{sharpness} 
that verifies the tightness of the density forecasting 
and the \textit{reliability}
that attests the distribution's statistical significance. For a detailed description see \citet{NowotarskiWeron2018, BavieraMessuti2019} and references therein. 
Let us briefly summarise their main characteristics.

\textit{Sharpness} is measured via the pinball loss function. 
It is an error measure for quantile forecasts. Pinball allows both a qualitative and a quantitative comparison of models.
In the former case, a symmetric pinball loss denotes 
that the density forecasting reflects with the same precision right and left tails of the true distribution. 
In the latter case, 
to quantitatively asses the sharpness we compute the Average of the Pinball Loss (APL) over 99 percentiles (from 1\% to 99\%); 
APL was the quantitative measure used to rank models in the GEFCom competition 2017 \citep[see, e.g.,][]{HONG2016896}.

\textit{Reliability} measures the consistency of the predicted distribution with the realised observations
and the robustness over time of the performance of the selected configuration of hyper-parameters.
The former is verified via a backtesting of the confidence intervals (hereafter CI);
e.g.,
if 90\% of the realised observations fall within the predicted 90\% CI, then this CI is considered reliable.
 From a quantitative point of view Unconditional Coverage tests this zero hypothesis.
 Moreover, Conditional Coverage tests the zero hypothesis that failures of the CI are not clustered together in the time-series. 

\noindent
The latter is checked, as described in the fourth step of Section \ref{subsec: NAX},
via the measure of some performance indicators both for point and for density forecasting over four years (2013-2016), 
keeping the optimal hyper-parameter configuration
chosen in the validation step.
In the following section we summarise the main results.
\section{Results}
\label{sec:results}	

In the previous section we have discussed that, for both model validation and testing, after data pre-processing,
the analysis is divided into three stages: calibration, forecasting and evaluation (see Figure \ref{fig: flowchart}).

The data pre-processing consists in the treatment of leap years and outliers.
We remove from the data-set February the $29^{th}$ in leap years,
while outliers are treated as in
\citet{Benth2008}:
in the 2007-2010
training set no outlier is detected. 

\begin{figure*}
			\centering
		\hspace{-1.2cm}
		\includegraphics[width=0.5\textwidth]{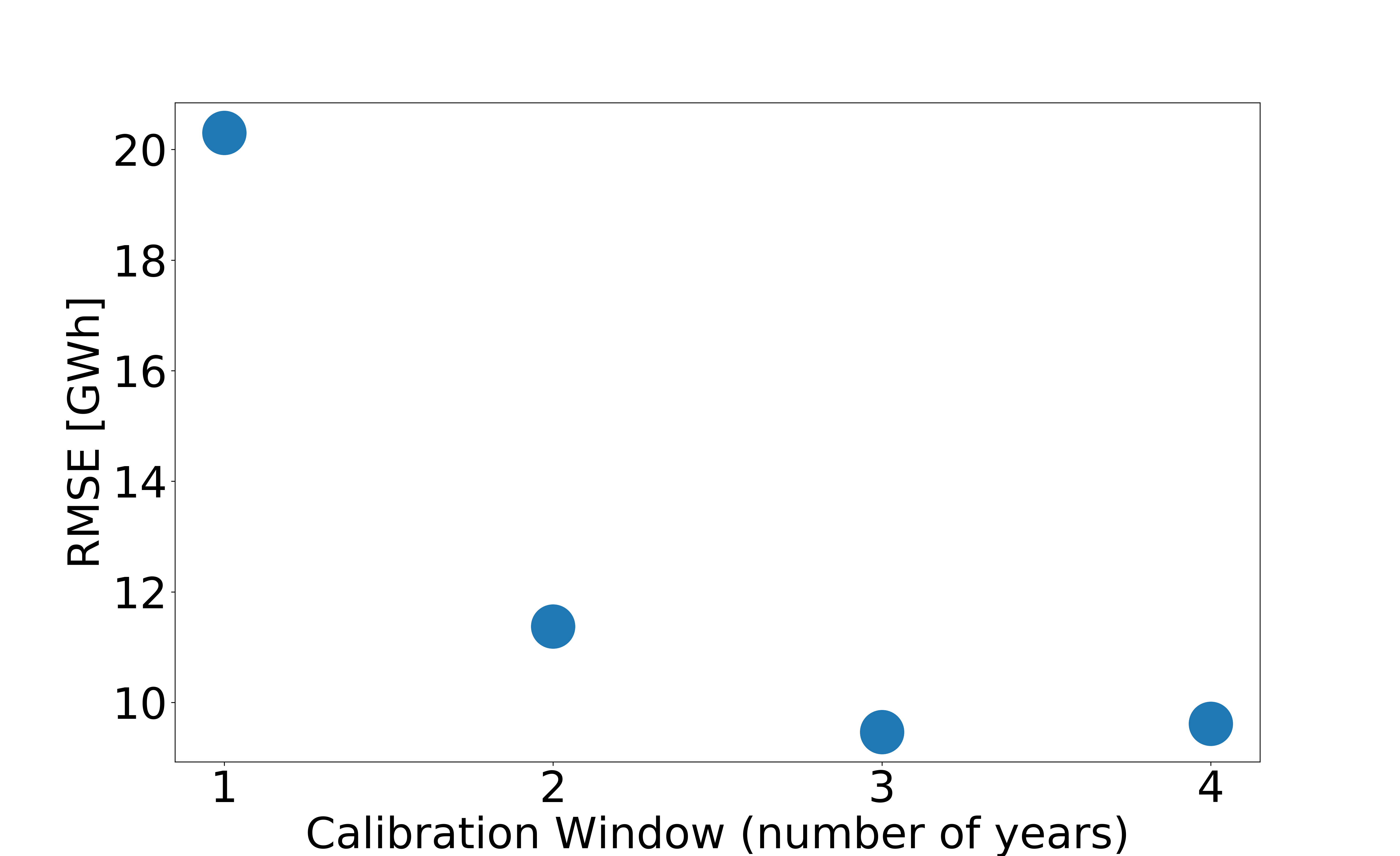}
		\caption[]{\small RMSE on the validation set (2011) with a training window of 1 (year $2010$), 2 (years $2009$ and $2010$), 3 (years $2008-2010$) and 4 years ($2007-2010$). For every time window we show the RMSE of the best among the calibrated models. We notice an improvement extending the training window up to three years. }
		\label{fig:hist}
\end{figure*}

For validation,
we train the NAX on every combination of the hyper-parameters reported in Table \ref{table:Hyperparam}. 
In particular, we select the training set time-window: 
as already mentioned in the previous section, 
this is a relevant hyper-parameter because the power market is rapidly evolving. It is crucial to train the NN on a time interval that helps the NN in its ability to generalise. Starting from one year, we consider longer intervals with yearly steps: we notice an improvement extending the training window up to three years, while for 
the four-year time window RMSE is higher.
In Figure \ref{fig:hist} we compare the RMSE on the validation set (2011) of the best selected model varying the length on the training window.
The best performing configuration on the validation set is the one with 3 neurons in the hidden layer, 
softmax activation, a three-year training window, 0.003 initial learning rate, a batch size of 50 observations and 
a regularization parameter equal to $10^{-4}$.
Moreover, we 
compare the NAX model with two simple benchmarks in the power industry: 
the GLM (the simplest benchmark model that does not include any weather information) and the ARX model \citep[see, e.g.,][p.534]{box15},
where an autoregressive (AR) component appears among linear regression covariates together with weather conditions and calendar effects (as eXogenous variables).
For comparison, we also consider another machine learning approach: 
a Gaussian Process with eXogenous inputs (GPX), see e.g. \citet[][]{BavieraMessuti2019}.
Among all analysed models the selected NAX 
is the best performer in the validation set.

For testing, 
the NAX model 
is calibrated on the new training set 
(from the $1^{st}$ of January $2009$ to the $31^{st}$ of December $2011$)
fitting first the GLM and then training the NN.
GLM parameters calibrated on the 2009-2011 time window are reported in \Cref{tab:GLMcoeff}. 
\begin{table}[h!] 
 \centering 
 {\footnotesize \begin{tabular}{|l|c|lr|} 
\hline 
& & Estimate & (SE) \quad \\
\hline 
Intercept & $\beta_0$ & 0.385*** & (0.023) \\ 
\hline 
Trend & $\beta_1$ & -0.16*** & (0.04) \\
\hline
$\sin\left(\omega t\right) $& $\beta_2$ & -0.003& (0.004) \\ 
\hline

$\cos\left(\omega t\right) $& $\beta_3$ & -0.028***& (0.004) \\ 
\hline

$\sin\left(2\omega t\right)$ & $\beta_4$ & 0.136*** & (0.004) \\ 
\hline

$\cos\left(2\omega t\right)$ & $\beta_5$ & -0.043*** & (0.004) \\ 
\hline

Sunday & $\beta_6$ & -0.146*** & (0.008) \\ 
\hline
Saturday & $\beta_7$ & -0.120*** & (0.008) \\ 
\hline

Holiday & $\beta_8$ & -0.060*** & (0.016) \\ 
\hline 
 \end{tabular}} 
 \caption{\small GLM parameters calibrated on the training data-set 2009-2011 with their standard deviation (SE). With *** we indicate
statistical significance of the parameters at 1\% significance level.} 
 \label{tab:GLMcoeff}
\end{table}



\begin{figure}
 \begin{subfigure}[b]{0.5\textwidth}
 \includegraphics[width=\textwidth]{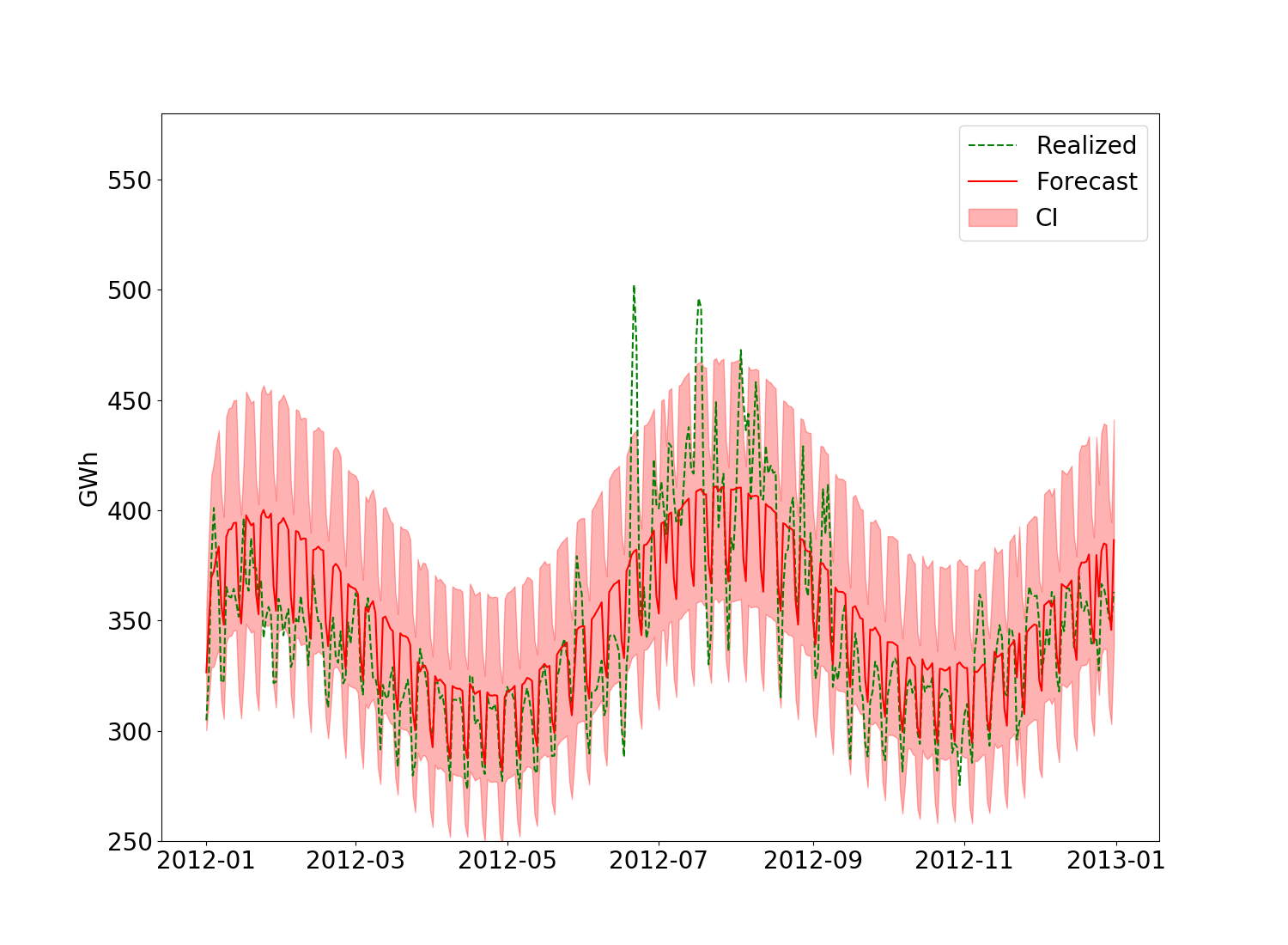}
 \end{subfigure}
 \begin{subfigure}[b]{0.5\textwidth}
 \includegraphics[width=\textwidth]{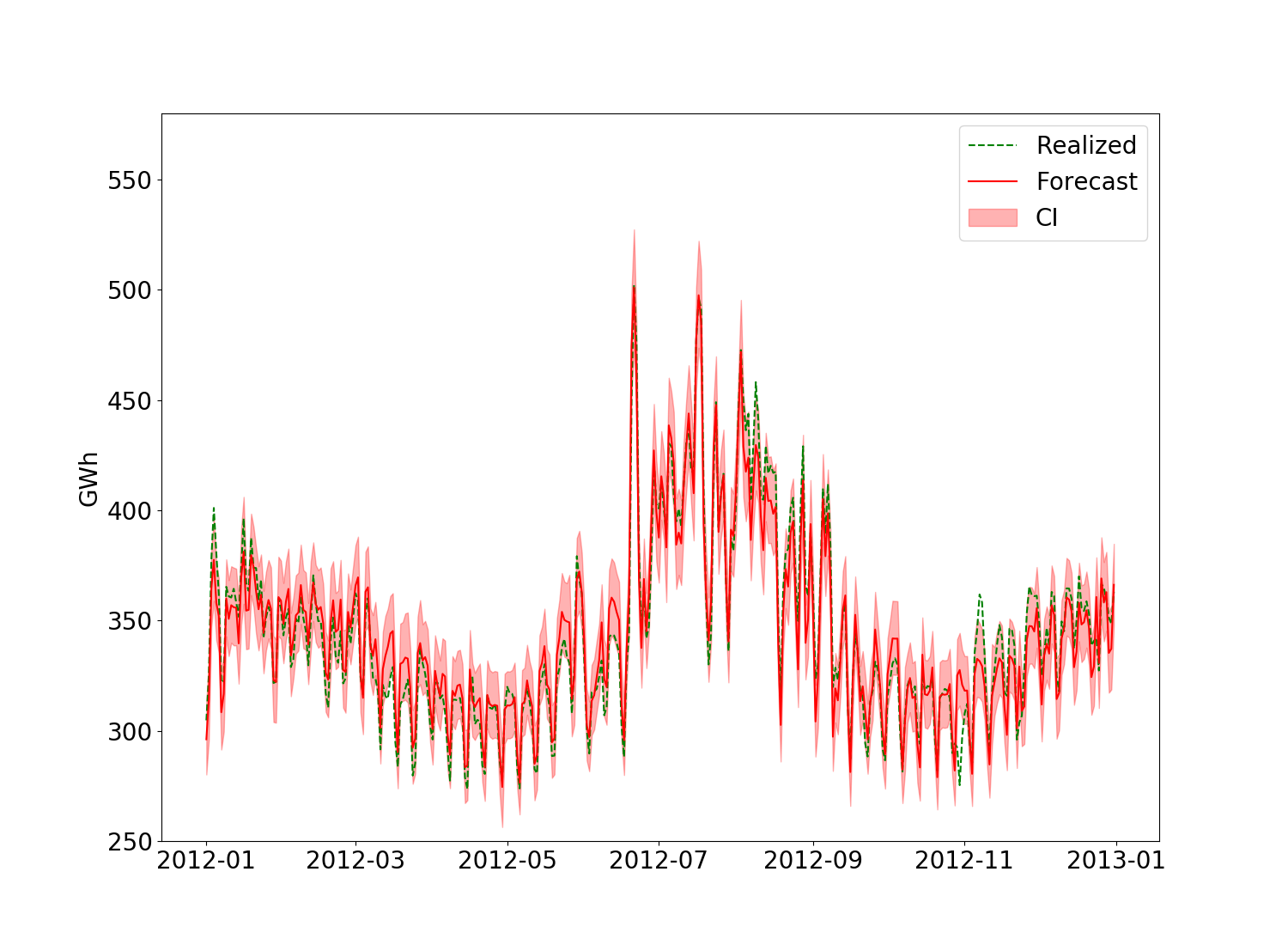}
 
 \end{subfigure}
 		\caption[]{\small ARX (on the left) and GPX (on the right) power consumption 
middle-term density prediction. 
Realised (dashed green line) and expected (red line) power consumption in GWh between January and December 2012 with predicted $95\%$ CI (transparent red).}
 		\label{fig:prendGAUARX}
\end{figure}

\begin{figure*}[h!]
		\centering
		\hspace{-1.2cm}
		\includegraphics[width=0.9\textwidth]{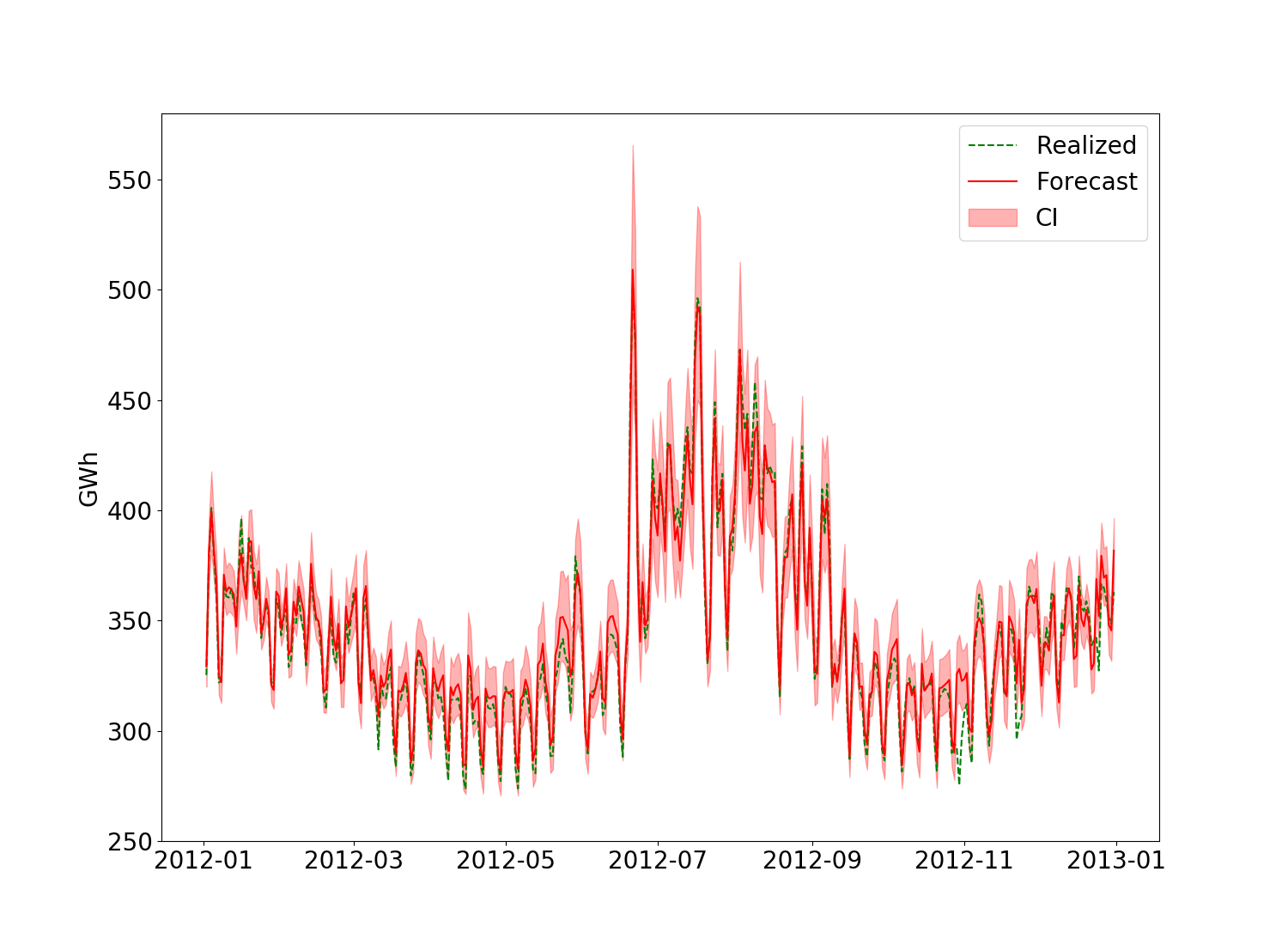}
	\caption[]{\small NAX power consumption middle-term density prediction. Realised (dashed green line) and expected (red line) power consumption in GWh between January and December 2012 with predicted $95\%$ CI (transparent red).
We observe that i) in terms of point forecasting, the model is able to provide a precise forecast even of spikes in consumption over summertime and
ii) in terms of density, realised consumption falls within the $95 \%$ CI in all but $15$ days ($95.8\%$ of cases). }
 \label{fig:powerpred}
\end{figure*}

 {\it Ex-post} forecasting is straightforward with the four models. We evaluate the model both qualitatively and quantitatively.
From a qualitative perspective
we plot models' forecast on the test set:
we consider ARX and GPX in Figure \ref{fig:prendGAUARX},
while in Figure \ref{fig:powerpred} we show the performance of NAX.
The continuous red line indicates the point forecasting 
while the transparent bright red indicates the 95\% CI; we also show with a dashed green line the power consumption realised on the test set. 
We observe that GPX improves significantly 
ARX performances both in terms of point and density forecasting,
but is less accurate than NAX. In particular, ARX CI look quite large.
As regards the NAX model, even if each density forecast at time
$t$ is a simple Gaussian with mean $T_t+S_t+\mu_t$ and variance
$\sigma_t^2$, in Figure \ref{fig:powerpred}
results look 
impressive:
not only the point forecast tracks closely the realised consumption (even the spikes in the summertime are tracked accurately),
but also the realised consumption falls within the $95\%$ CI in all but 15 days (95.8 \%), 
and the densities reproduce the observed behaviour of 
periods of high volatility in summertime followed by periods of low volatility in 
wintertime.\footnote{Let us underline that the last power consumption considered in
model calibration is the $31^{st}$ of December $2011$, while the
forecasting goes up to one year later to the $31^{st}$ of December $2012$.}

In the remaining part of this section we discuss some quantitative criteria that show the goodness
of both the point and the density forecasting comparing the results with the ones of GLM, ARX and GPX.
We first consider accuracy measures for the point forecasting and then we show the results for sharpness and reliability of the probabilistic forecasting:
these evaluation techniques have been described in Section \ref{sec:methodology}.
	

\begin{table}[h!!]
		\centering
		{\footnotesize \begin{tabular}{|c|c|c|c|c|c|} 
			\hline 
			& GLM & ARX & GPX&\textbf{NAX} \\ 
			\hline 
			RMSE [GWh]& 26.69 &26.13 & 10.74& \textbf{8.10} \\ 
			\hline 
			MAPE (\%) & 6.00 & 5.80 &2.50& \textbf{1.74} \\ 
			\hline 
		\end{tabular}}
\caption{\small RMSE and MAPE for the four models considered on the test set (2012). 
We observe that NAX not only presents a MAPE lower than $5\%$, 
i.e. it is considered a good forecast by practitioners, 
but also the lower absolute error (RMSE) indicates a more precise point forecasting in summertime, i.e. when forecasting is more relevant due to the spikes in consumption.}
		\label{tab:sharp_errors}
\end{table}

First, we compute RMSE and MAPE 	
of the point forecasting on the test set (see Table \ref{tab:sharp_errors}).
One can notice that GPX and NAX are significantly better than GLM and ARX in terms of RMSE and MAPE, being NAX the best one. 
In particular, both GPX and NAX are below the $5\%$ threshold that 
characterises good power forecasting models for practitioners.
Moreover,
a lower RMSE (more than two-thirds lower than GLM and almost a third lower than GPX) indicates that NAX reduces significantly the error also in summertime, 
when the forecasting is more relevant due to higher consumption and volatility: a behaviour we have observed in Figure \ref{fig:prendGAUARX} and Figure \ref{fig:powerpred}.

\bigskip
		
\begin{figure*}[h!]
	\centering
	\includegraphics[width=0.9\textwidth]{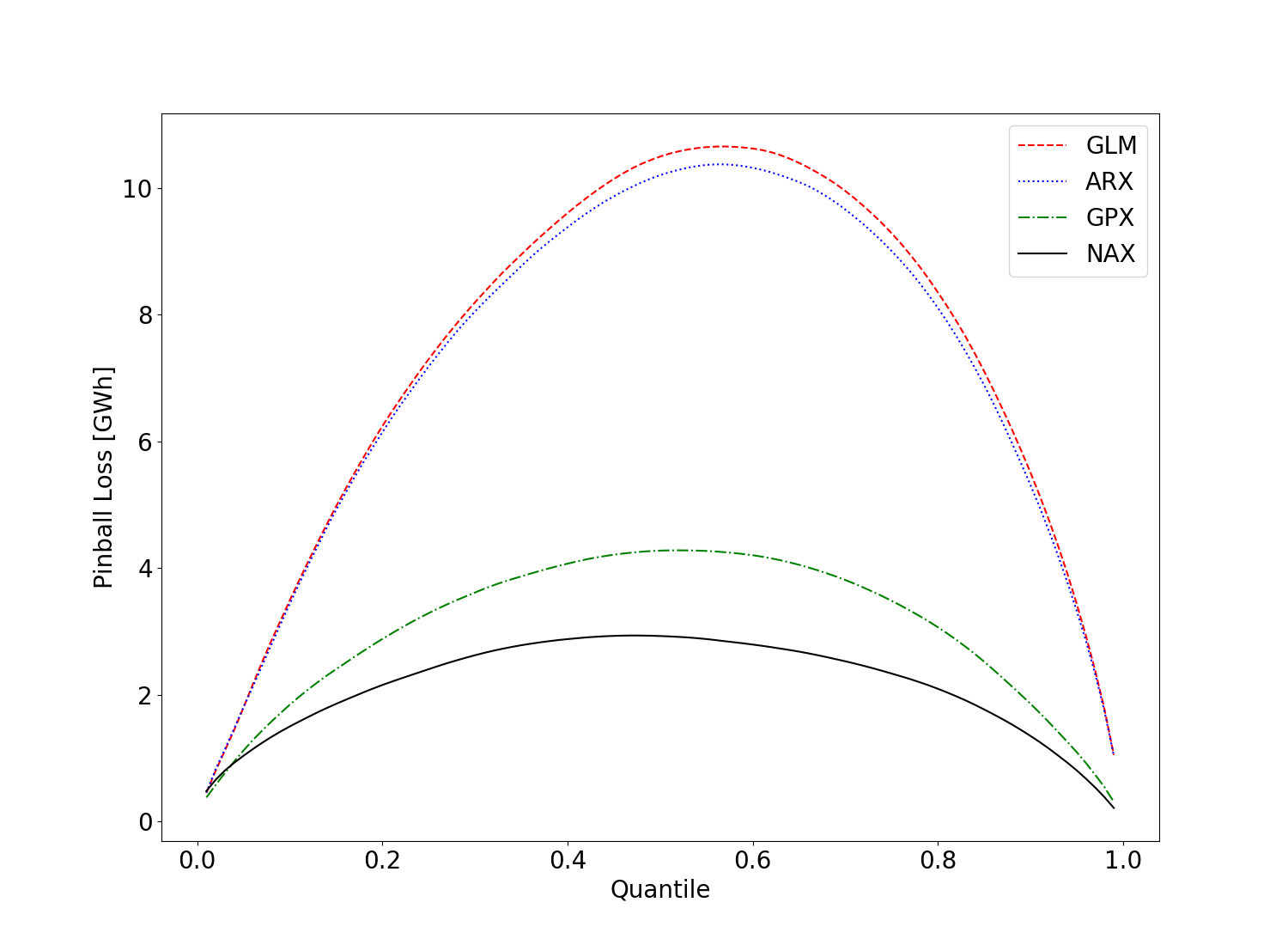}
	\caption{\small Pinball loss functions for the four models for every percentile. 
We observe that not only NAX 
presents the lowest score for all percentiles (i.e. it is sharper and more accurate), 
but also that the pinball loss is symmetric for both NAX and GPX.}
	\label{fig:pinball}
\end{figure*}

Second, we consider the analysis of sharpness.
 Figure \ref{fig:pinball} represents the pinball loss 
for the four models on every percentile. 
We observe that for all percentiles (with one exception) the NAX pinball loss is significantly lower 
than the other three models' pinball loss.
The plot of the pinball loss provides useful information 	
not only in terms of sharpness 
but also related to its symmetric shape:
density forecasting of power consumption,
for both NAX and GPX,
is reproducing with 
a similar accuracy both right and left tails of the actual consumption density. NAX shows to be the most accurate, being the pinball loss the lowest.
	

\bigskip

\begin{figure*}[h!]
	\centering
	\includegraphics[width=0.9\textwidth]{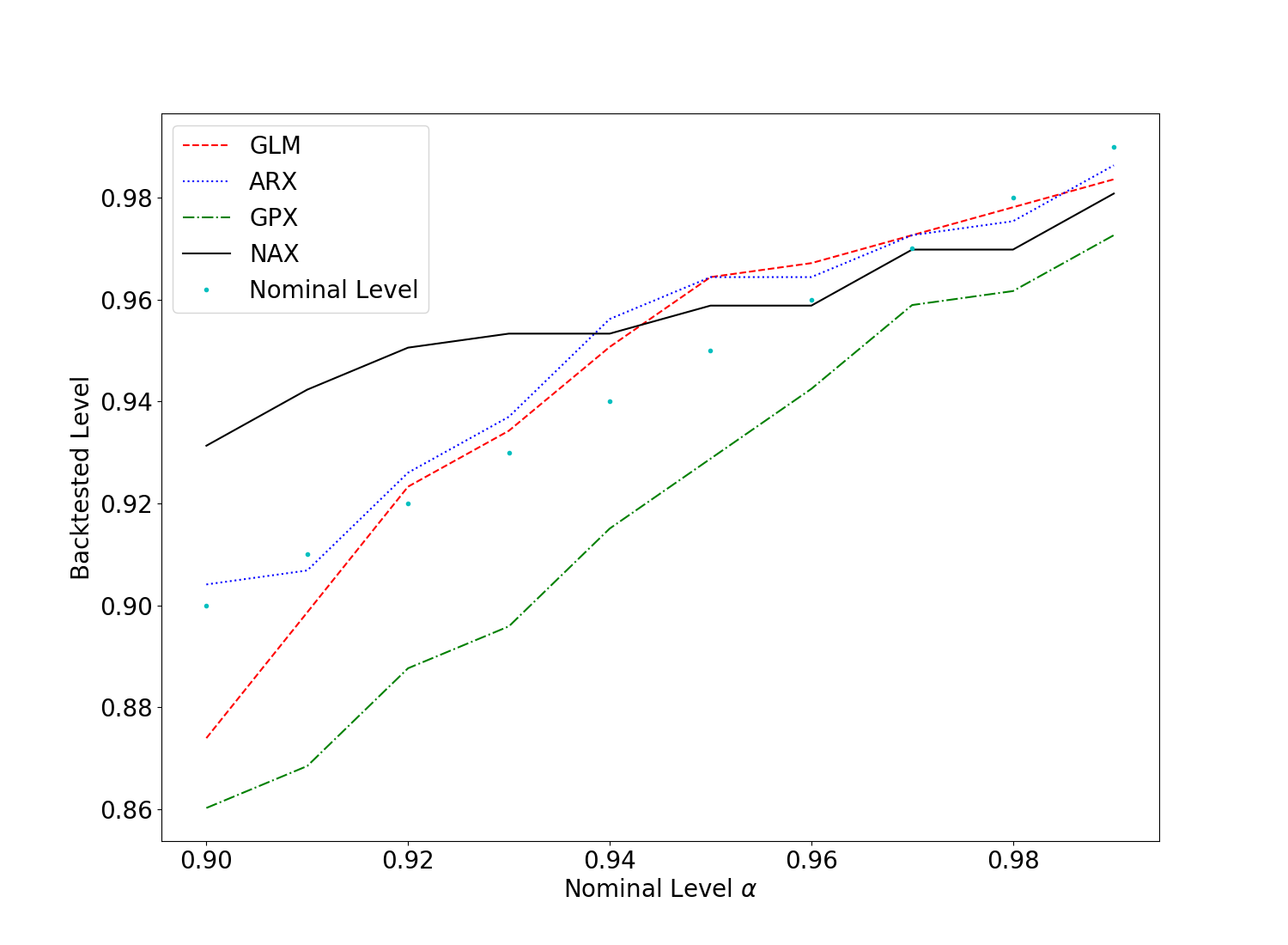}
	\caption{\small Backtested CI. 
We show the fraction of times the realised power consumption falls inside a 
confidence level $\alpha$ {\it vs} $\alpha$. 
We observe that the empirical coverage for GPX and NAX are very close to the nominal levels for high quantiles, being NAX significantly better than GPX.}
	\label{fig:backtest}
\end{figure*}

Finally, we test model robustness both backtesting the CI 
and verifying its reliability over-time. 
First, we backtest CI, counting the fraction of days the realised power consumption falls outside a 
confidence level $\alpha$ of the predicted density (violations) and comparing to the nominal ones, equal to $1 - \alpha$. 
One can observe in Figure \ref{fig:backtest} that the NAX backtested CI are close to the nominal values. 
GLM and ARX backtested quantiles look better but 
this is due to the fact that CI are much larger:
one must take into account the sharpness of NAX and GPX CI compared to the linear benchmarks (see Figures \ref{fig:prendGAUARX}
and \ref{fig:powerpred}). In any case, both GPX and NAX provide good results.

These results on CI are 
tested also from a quantitative point of view through likelihood ratio tests. 
Table \ref{tab:likelihoodratios} resumes 
the results of these 
tests, standard in the backtesting of VaR in the banking industry after the introduction of Basel II: the Unconditional Coverage and the Conditional Coverage. The Unconditional Coverage is a test on the number of violations while the Conditional Coverage tests whether violations tend to cluster in some particular periods.
All models pass the Unconditional Coverage test while no model passes the Conditional Coverage. 
This result suggests to analyze more in detail the time-series of violations; one can observe that for the NAX model violations cluster only in 
autumn and spring time (and this causes the test failure), while in winter and summer, i.e. the most interesting periods for power forecasting,
violations are distributed regularly.\footnote{Analysis available upon request.}

\begin{table}[h!]
		\centering
		{\footnotesize 	\begin{tabular}{|c|c|c|c|c|c|} 
			\hline
			& GLM & ARX &GPX & \textbf{NAX} & $\chi^2$ Statistic \\ \hline
			Unconditional Coverage &1.76 & 1.76 & 3.08 & \textbf{0.63} & 3.84 \\ \hline 
			Conditional Coverage &27.28 & 27.28&13.86 &\textbf{22.84} &5.99 \\ \hline 
		\end{tabular}}
		\caption[Likelihood Ratios]{\small Likelihood Ratios tests at 95\% level of 95\% CI. The $\chi^2$ test statistic represents the threshold over which one should reject the null hypothesis. }
		\label{tab:likelihoodratios}
\end{table}

Then,
to verify the robustness over time of the NAX and the GPX performances, we evaluate the
forecasts on the years from 2013 to 2016; 
we consider a NAX model with the same hyper-parameters selected in the 
validation step and 
we calibrate only model parameters on the 
new training time-window (e.g. $2010-2012$ for the year $2013$).
This analysis is particularly relevant for machine learning techniques that involve several hyper-parameters.
In Table \ref{tab:robustnes_testing} we compare
MAPE, RMSE and APL of GPX and NAX on the different test sets (from 2012 to
2016). In every year MAPE falls well below the 5\% threshold; 
NAX always performs better than GPX (except for the 2015
testing year) in terms of both point and density forecasting.\footnote{Notice the differences
between the two modelling approaches. While for GPX there is a
standard model calibration of model parameters on the three-year
calibration period, for NAX we are considering the same
hyper-parameters selected in the validation and we are just
training model weights.}

\begin{table}[h!]
 \centering
 {\footnotesize \begin{tabular}{|c||c|c||c|c||c|c|} 
 \hline 
 & \multicolumn{2}{c||}{MAPE (\%)} & \multicolumn{2}{c||}{RMSE [GWh]} & \multicolumn{2}{c|}{APL [GWh]}\\ \hline
 &GPX&\textbf{NAX} &GPX&\textbf{NAX} &GPX&\textbf{NAX} \\ \hline 
 2012 & 2.50& \textbf{1.74}& 10.74& \textbf{8.10}& 3.04& \textbf{2.15} \\ \hline 
 2013 &2.17& \textbf{2.13} &10.30& \textbf{10.10} &2.82& \textbf{2.76} \\ \hline 
 2014 &2.39& \textbf{2.00} &10.66& \textbf{8.70} &2.95& \textbf{2.49} \\ \hline 
 2015 &2.41& \textbf{2.52} &10.31& \textbf{10.74} &2.93& \textbf{3.05} \\ \hline 
 2016 &3.16& \textbf{1.70} &13.46& \textbf{7.58} &3.94& \textbf{2.07} \\ \hline 
 \end{tabular}} 
\caption{\small MAPE, RMSE and APL for GPX and NAX on different test sets (from 2012 to 2016). 
We observe that NAX and GPX in all testing year presents a MAPE lower than $5\%$. The MAPE values are stable and NAX has always the best MAPE, RMSE and APL except that in the 2015 testing year. }
\label{tab:robustnes_testing}

\end{table}

Summarizing, 
Table
 \ref{tab:robustnes_testing} together with Figure
 \ref{fig:backtest} are the strongest results of our analysis. 
On the one hand, 
Figure \ref{fig:backtest} shows that the proposed model reproduces
accurately observed tail distribution with a frequency of violations
close to the
nominal level:
not only we have a MAPE lower than 5\% over the whole time-horizon, 
but the nominal level and the empirical coverage appear close for all values of $\alpha$. 
On the other hand,
Table \ref{tab:robustnes_testing} shows that
the NAX model can provide 
robust over time point and density forecasts for daily power consumption, 
even without changing the optimal model hyper-parameters. 

\section{Ex-ante prediction}
\label{sec:exante}

In the previous section we have evaluated the model density forecasting performance using realised temperature data 
({\it ex-post} forecasting): as already discussed, this is the most relevant piece of information for both academicians and practitioners, because it describes the quality of the model relation between consumption and temperatures --even on a middle term horizon-- without embedding forecasting errors in temperatures.

One possible application is to verify also the {\it ex-ante} prediction power of NAX in a real-world situation simulating the daily temperatures on the testing set 
(2012).
An analytical formula for the Cumulative Distribution Function (CDF) for the {\it ex-ante} residual $R_t$ 
can be obtained conditioning on the temperature time-series. 
For every $K\in \mathbb{R}$ 
\begin{align*}
 \mathbb{P}\left(R_t<K\right)=
  \mathbb{E}\left[ \mathbb{P}\left(R_t < K |\textbf{T}\right)\right]=
  \mathbb{E}\left[N\left(\frac{K-\mu_t(\textbf{T})}{\sigma_t(\textbf{T})}\right)\right]\;\;,
\end{align*}
where $\textbf{T}$ is the random time-series of daily temperature,
$\mathbb{E}\left[ \bullet \right]$ is the expectation w.r.t. the simulated temperature $\textbf{T}$ and $N\left(\bullet \right)$ is the CDF of a standard normal r.v..
NAX provides the conditional CDF given the time-series $\textbf{T}$:
$\mu_t(\textbf{T})$ and $\sigma_t(\textbf{T})$ are the NAX outputs computed considering as imputs the temperatures $\textbf{T}$. 
We can estimate the cumulative distribution by giving the same probability to every simulated temperature time-series. 

For the temperature simulation we follow the approach introduced by \citet[][pp.7-8]{HyndmanFan2010}. 
One bootstrapped temperature sequence consists of a sample of blocks of length $L$ randomly selected with a uniform distribution in ($m-\Delta$, $m+\Delta$). 
The block comes from a different (previous) randomly selected year (between 2007 and 2011)
and a random shift $S$ uniform in $(-\Delta,$ $\Delta)$ is applied on the days of the year. 
We select $m=7$ and $\Delta=3$ and we simulate 2000 paths as in \citet[][p.9]{HyndmanFan2010}.
For example, we start the bootstrapped time-series from the $1^{st}$ of January with a block 1 of 7 days; this block may come from 2010, block 2 can have a $L=9$ and $S=1$ 
the 9 day block may come from the period starting on the $9^{th}$ of January 2009, and so on up to the $31^{st}$ of December.    
Differently from \citet{HyndmanFan2010}, we do not add any additional noise to the temperatures to closely match those observed in the time window we consider.

\bigskip

We can then evaluate the quality of the obtained {\it ex-ante} forecast.

First, we compare the RMSE and MAPE of {\it ex-post} and {\it ex-ante} point forecasting in Table \ref{tab:sharp_errors_ex_ante}. 
As expected, we observe an increase in the point forecasting error.

\begin{figure*}[h!!]
		\centering
		\hspace{-1.2cm}
		\includegraphics[width=0.9\textwidth]{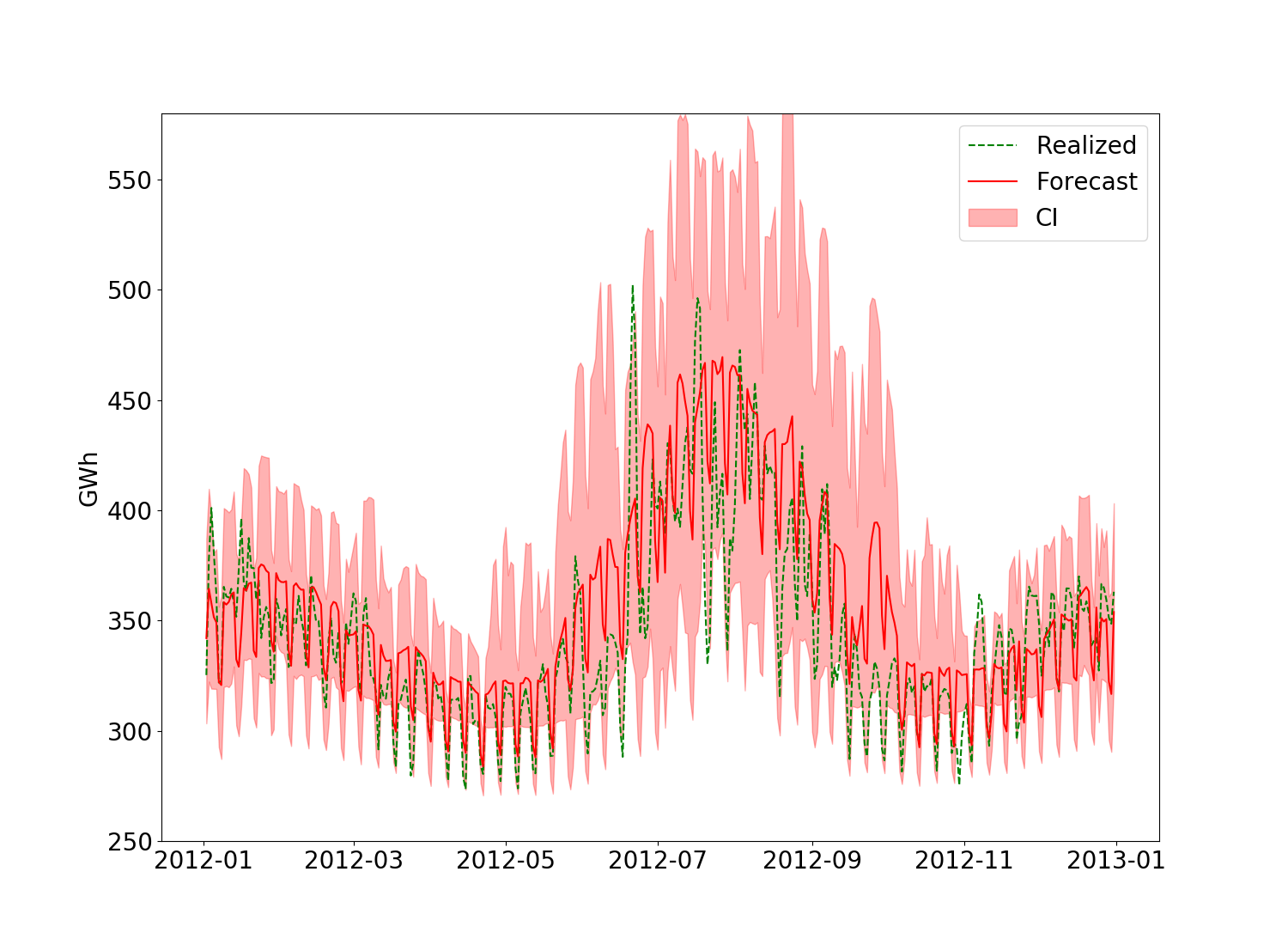}
	\caption[]{\small  NAX power consumption {\it ex-ante} middle-term density prediction.
 Realised (dashed green line) and expected (red line) power consumption in GWh between January and December 2012 with predicted $95\%$ CI (transparent red).
We observe that i) in terms of {\it ex-ante} point forecasting, the model can provide a valid forecast but it misses some spikes in summertime and wintertime
ii) in terms of density, realised consumption falls within the $95 \%$ CI in all but $22$ days ($94.0\%$ of cases). The CI appear wider than in the {\it ex-post} case (cf. Figure \ref{fig:powerpred}).}
 \label{fig:powerpred_exante}
\end{figure*}

\begin{table}[h!!]
		\centering
		{\footnotesize \begin{tabular}{|c|c|c|c|c|c|} 
			\hline 
			& {\it ex-ante} & {\it ex-post} \\ 
			\hline 
			RMSE [GWh]& 28.99 &{8.10} \\ 
			\hline 
			MAPE (\%) & 6.01&{1.74} \\ 
			\hline 
		\end{tabular}}
\caption{\small RMSE and MAPE for {\it ex-ante} and {\it ex-post} forecast on the test set (2012). 
As expected we observe higher errors for the {\it ex-ante} forecast because we are considering the uncertainty on temperatures.
}
\label{tab:sharp_errors_ex_ante}
\end{table}
Second, in Figure \ref{fig:powerpred_exante} we show NAX power consumption middle-term {\it ex-ante} density prediction. 
As in Figure \ref{fig:powerpred} for the {\it ex-post} case, 
the continuous red line indicates the point forecasting 
while the transparent bright red indicates the 95\% CI; we also show with a dashed green line the power consumption realised on the test set. 
As expected, results are less accurate than in the {\it ex-post} prediction both in terms of point forecasting and of CI wideness because we are considering the error in temperature forecasts. Despite this, 
the seasonality in both the volatility and the point forecasting is preserved, and
the reliability of CI remains high (realised consumption falls within the $95 \%$ CI in all but $22$ days, $94.0\%$ of cases).

\begin{figure*}[h!]
		\centering
		\hspace{-1.2cm}
		\includegraphics[width=0.85\textwidth]{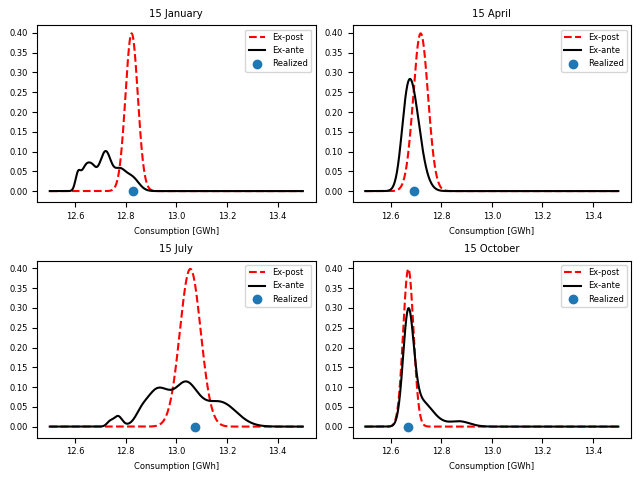}
	\caption[]{\small 
{\it Ex-post} (continuous black line) and {\it ex-ante} (dashed red) forecasted densities  for the $15^{th}$ of January, April, July and October 2012. 
In all cases, we observe a wider distribution in the {\it ex-ante} forecasting and the realised consumption falls within the densities. 
The seasonality of volatility is preserved.
Notice that the {\it ex-ante} distributions are significantly wider in summer and winter, 
 while during spring and autumn the two densities appear quite close.}.
 \label{fig:distro}
\end{figure*}

To show
the forecasting density change after introducing the error in temperatures, in Figure \ref{fig:distro} 
we analyze four days (in winter, spring, summer and autumn). In all cases, we observe a wider distribution in the {\it ex-ante} forecasting and the realised consumption falls within the densities. Notice that the {\it ex-ante} distributions are always wider, in particular in summer and winter, while during spring and autumn the two forecasts look quite close.

\begin{figure*}[h!]
	\centering
	\includegraphics[width=0.9\textwidth]{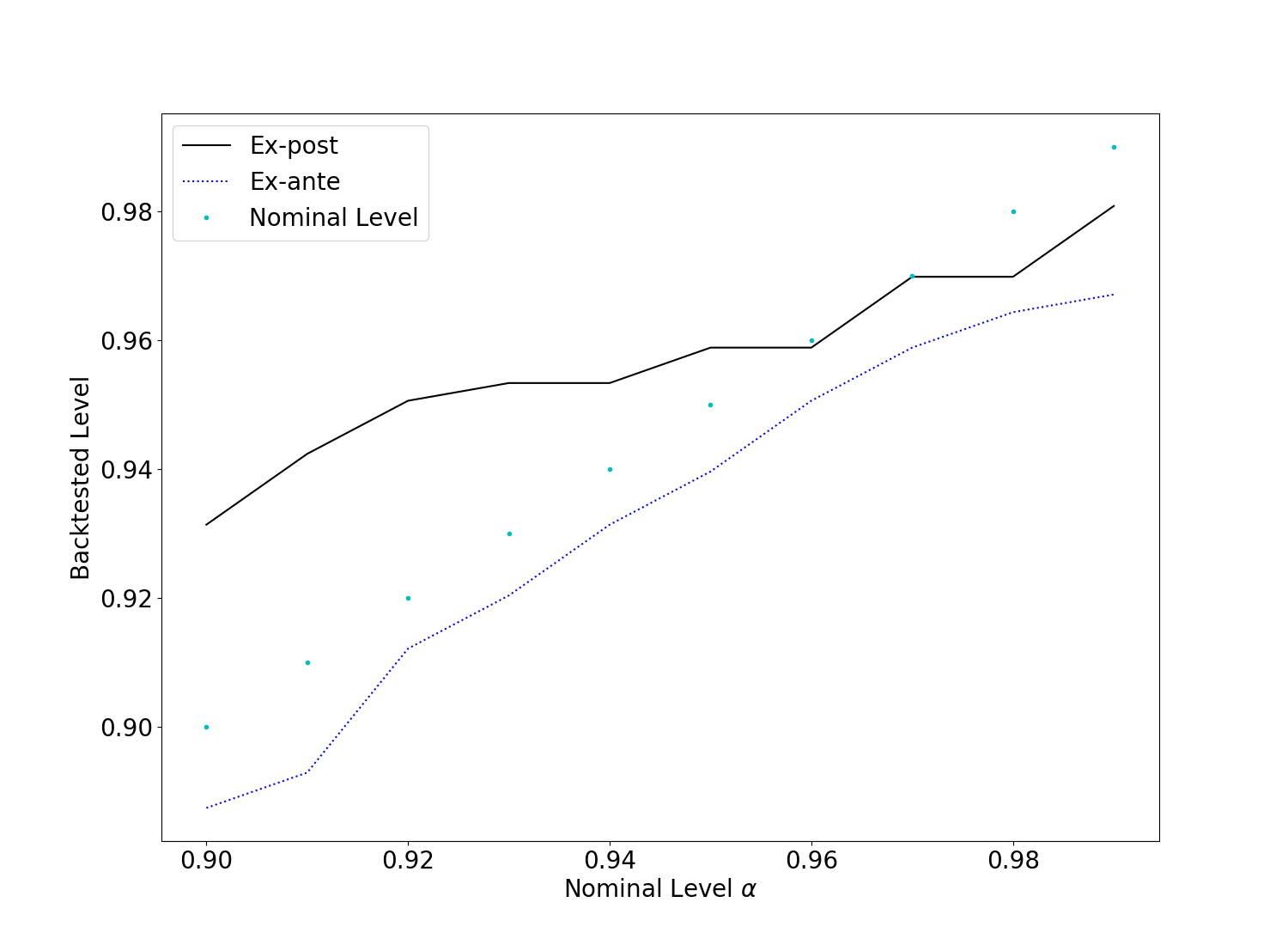}
	\caption{\small Backtested CI. 
We show the fraction of times the realised power consumption falls inside a predicted
confidence level $\alpha$ {\it vs} $\alpha$. 
We observe that the empirical coverage for {\it ex-ante} and {\it ex-post} prediction are close to the nominal levels for high quantiles.}
	\label{fig:backtest_ex_ante}
\end{figure*}

Finally, we can backtest CI, counting the fraction of days the realised power consumption falls outside a 
confidence level $\alpha$ of the predicted density (violations) and comparing to the nominal values, equal to $1 - \alpha$. 
One can observe in Figure \ref{fig:backtest_ex_ante} that both {\it ex-post} and {\it ex-ante} CI are close to the nominal values. 
One must take into account the sharpness of  {\it ex-post} CI compared to the {\it ex-ante} CI (cf. Figures \ref{fig:powerpred}
and \ref{fig:powerpred_exante}).

\bigskip

In this section we have presented the {\it ex-ante} density forecast. 
As expected, we observe an increase in the point forecasting error and a widening of the CI w.r.t. the {\it ex-post} case. 
However, the density forecasting remains sufficiently accurate and the seasonality structure (both in the point forecasting and in the volatility) is preserved,
showing the robustness of the NAX density forecasts.

\section{Concluding remarks}
\label{sec:conclusions}
In this paper we have introduced the NAX model 
for middle-term density forecast of power consumption, a hybrid model
that joints the advantages of
classical univariate time-series analysis and a shallow NN.

The main contributions of this paper are threefold.
the NN architecture --described in Figure \ref{fig:nn_scheme}-- is new in the literature: 
it is designed to obtain both density forecasting and the autoregressive feature observed in the consumption time-series. 
Second, we have shown that NAX achieves an excellent middle-term  
forecasting of power consumption using only weather data.
We achieve an {\it ex-post} MAPE lower than 2\% on the testing set (cf. Table \ref{tab:sharp_errors}), 
a $20\%$ better than GPX in relative terms, and realised power consumption is forecasted with great accuracy (cf. Figure \ref{fig:powerpred}). 
Third, we have valued the density forecasting via some sharpness and reliability measures, 
showing the quality of the achieved results (cf. Figure \ref{fig:pinball} and \ref{fig:backtest}). 
The robustness over time of NAX performances, 
both in terms of point (MAPE and RMSE) and density (APL) forecasting, 
has been tested from 2013 to 2016 
(cf. Table \ref{tab:robustnes_testing}).
We have also shown an {\it ex-ante} middle-term density forecasting as an application of the proposed technique.

These results show that the proposed model 
can catch accurately the relation between weather conditions and power consumption distribution in a sharp and reliable way,
even considering small time-series with only a few years long training window of daily data. 

\section*{Acknowledgements}
We thank Y. Chen, M. De Prato, S. Grassi, J. Turiel and all participants to the workshop 
 {\it ``AI, Financial Automation and Market Risk''} (2020). 
This work was supported by  {\it ``Ho2020 A FINancial supervision and TECHnology compliance training programme''}.
\section*{Abbreviations}
\addcontentsline{toc}{chapter}{Abbreviations}
\markboth{}{Abbreviations}

{\small \begin{tabular}{lll}
		$AR$ & : & Auto Regressive model 	\\
		$ARX$ & : & Auto Regressive eXogenous model 	\\
		$APL$ & : & Average Pinball Loss \\ 
		$CI$ & : & Confidence Interval 	\\
		\text{cf.} & : & \text{compare; from Latin: confer} \\
		$GLM$ & : & General Linear Model 	\\
		$GPX$ & : & Gaussian Process eXogenous model 	\\
		$IS$ & : & In-Sample (also training set) 	\\
		$MAPE$ & : & Mean Absolute Percentage Error \\
		$NAX$ & : & NN with Autoregression and eXternal inputs: the proposed model 	\\
		$NN$ & : & Neural Network 	\\
		$OS$ & : & Out-of-Sample 	\\
		$RMSE$ & : & Root Mean Squared Error \\
		$SE$ & : & Standard Error \\
		w.r.t. & : & with respect to \\
\end{tabular}}

\bigskip

\bibliography{references}

\begin{thebibliography}{24}
\expandafter\ifx\csname natexlab\endcsname\relax\def\natexlab#1{#1}\fi
\providecommand{\url}[1]{\texttt{#1}}
\providecommand{\href}[2]{#2}
\providecommand{\path}[1]{#1}
\providecommand{\DOIprefix}{doi:}
\providecommand{\ArXivprefix}{arXiv:}
\providecommand{\URLprefix}{URL: }
\providecommand{\Pubmedprefix}{pmid:}
\providecommand{\doi}[1]{\href{http://dx.doi.org/#1}{\path{#1}}}
\providecommand{\Pubmed}[1]{\href{pmid:#1}{\path{#1}}}
\providecommand{\bibinfo}[2]{#2}
\ifx\xfnm\relax \def\xfnm[#1]{\unskip,\space#1}\fi
\bibitem[{Baviera \& Messuti(2020)}]{BavieraMessuti2019}
\bibinfo{author}{Baviera, R.}, \& \bibinfo{author}{Messuti, G.}
  (\bibinfo{year}{2020}).
\newblock \bibinfo{title}{Daily middle-term probabilistic forecasting of power
  consumption in {N}orth-{E}ast {E}ngland}.
\newblock {\it \bibinfo{journal}{Available online:
  arxiv.org/abs/2005.13005}\/},  (pp. \bibinfo{pages}{1--26}).
\bibitem[{Benth et~al.(2008)Benth, Benth \& Koekebakker}]{Benth2008}
\bibinfo{author}{Benth, F.~E.}, \bibinfo{author}{Benth, J.~S.}, \&
  \bibinfo{author}{Koekebakker, S.} (\bibinfo{year}{2008}).
\newblock {\it \bibinfo{title}{Stochastic modelling of electricity and related
  markets}\/}.
\newblock \bibinfo{publisher}{World Scientific}.
\bibitem[{Box et~al.(2015)Box, Jenkins, Reinsel \& Ljung}]{box15}
\bibinfo{author}{Box, G.~E.}, \bibinfo{author}{Jenkins, G.~M.},
  \bibinfo{author}{Reinsel, G.~C.}, \& \bibinfo{author}{Ljung, G.~M.}
  (\bibinfo{year}{2015}).
\newblock {\it \bibinfo{title}{Time series analysis: forecasting and
  control}\/}.
\newblock \bibinfo{publisher}{John Wiley \& Sons}.
\bibitem[{Chollet(2015)}]{chollet2015keras}
\bibinfo{author}{Chollet, F.} (\bibinfo{year}{2015}).
\newblock \bibinfo{title}{Keras}.
\newblock \bibinfo{howpublished}{\url{https://keras.io}}.
\bibitem[{Christoffersen(1998)}]{Christoffersen1998}
\bibinfo{author}{Christoffersen, P.~F.} (\bibinfo{year}{1998}).
\newblock \bibinfo{title}{Evaluating interval forecasts}.
\newblock {\it \bibinfo{journal}{International Economic Review}\/},  {\it
  \bibinfo{volume}{39}\/}, \bibinfo{pages}{841--862}.
\bibitem[{Fan \& Chen(2006)}]{FanChen2006}
\bibinfo{author}{Fan, S.}, \& \bibinfo{author}{Chen, L.}
  (\bibinfo{year}{2006}).
\newblock \bibinfo{title}{Short-term load forecasting based on an adaptive
  hybrid method}.
\newblock {\it \bibinfo{journal}{IEEE Transactions on Power Systems}\/},  {\it
  \bibinfo{volume}{21}\/}, \bibinfo{pages}{392--401}.
\bibitem[{Felder et~al.(2010)Felder, Kaifel \& Graves}]{FKG2010wind}
\bibinfo{author}{Felder, M.}, \bibinfo{author}{Kaifel, A.}, \&
  \bibinfo{author}{Graves, A.} (\bibinfo{year}{2010}).
\newblock \bibinfo{title}{Wind power prediction using mixture density recurrent
  neural networks}.
\newblock In {\it \bibinfo{booktitle}{European Wind Energy Conference \&
  Exhibition (EWEC: Warsaw, Poland, 20-23 April, 2010)}\/} (pp.
  \bibinfo{pages}{3417--3424}).
\bibitem[{Goude et~al.(2013)Goude, Nedellec \& Kong}]{GNK2014}
\bibinfo{author}{Goude, Y.}, \bibinfo{author}{Nedellec, R.}, \&
  \bibinfo{author}{Kong, N.} (\bibinfo{year}{2013}).
\newblock \bibinfo{title}{Local short and middle term electricity load
  forecasting with semi-parametric additive models}.
\newblock {\it \bibinfo{journal}{IEEE transactions on smart grid}\/},  {\it
  \bibinfo{volume}{5 , Issue: 1}\/}, \bibinfo{pages}{440 -- 446}.
\bibitem[{Hong \& Fan(2016)}]{HongFan2016}
\bibinfo{author}{Hong, T.}, \& \bibinfo{author}{Fan, S.}
  (\bibinfo{year}{2016}).
\newblock \bibinfo{title}{Probabilistic electric load forecasting: A tutorial
  review}.
\newblock {\it \bibinfo{journal}{International Journal of Forecasting}\/},
  {\it \bibinfo{volume}{32}\/}, \bibinfo{pages}{914--938}.
\bibitem[{Hong et~al.(2016)Hong, Pinson, Fan, Zareipour, Troccoli \&
  Hyndman}]{HONG2016896}
\bibinfo{author}{Hong, T.}, \bibinfo{author}{Pinson, P.}, \bibinfo{author}{Fan,
  S.}, \bibinfo{author}{Zareipour, H.}, \bibinfo{author}{Troccoli, A.}, \&
  \bibinfo{author}{Hyndman, R.~J.} (\bibinfo{year}{2016}).
\newblock \bibinfo{title}{Probabilistic energy forecasting: Global energy
  forecasting competition 2014 and beyond}.
\newblock {\it \bibinfo{journal}{International Journal of Forecasting}\/},
  {\it \bibinfo{volume}{32}\/}, \bibinfo{pages}{896 -- 913}.
\bibitem[{Hyndman \& Fan(2010)}]{HyndmanFan2010}
\bibinfo{author}{Hyndman, R.}, \& \bibinfo{author}{Fan, S.}
  (\bibinfo{year}{2010}).
\newblock \bibinfo{title}{Density forecasting for long-term peak electricity
  demand}.
\newblock {\it \bibinfo{journal}{Power Systems, IEEE Transactions on}\/},  {\it
  \bibinfo{volume}{25}\/}, \bibinfo{pages}{1142 -- 1153}.
\bibitem[{Kong et~al.(2017)Kong, Dong, Jia, Hill, Xu \& Zhang}]{Kong2017}
\bibinfo{author}{Kong, W.}, \bibinfo{author}{Dong, Z.~Y.},
  \bibinfo{author}{Jia, Y.}, \bibinfo{author}{Hill, D.~J.},
  \bibinfo{author}{Xu, Y.}, \& \bibinfo{author}{Zhang, Y.}
  (\bibinfo{year}{2017}).
\newblock \bibinfo{title}{Short-term residential load forecasting based on lstm
  recurrent neural network}.
\newblock {\it \bibinfo{journal}{IEEE Transactions on Smart Grid}\/},  {\it
  \bibinfo{volume}{10}\/}, \bibinfo{pages}{841--851}.
\bibitem[{Kupiec(1995)}]{Kupiec1995}
\bibinfo{author}{Kupiec, P.} (\bibinfo{year}{1995}).
\newblock \bibinfo{title}{Techniques for verifying the accuracy of risk
  measurement models}.
\newblock {\it \bibinfo{journal}{The Journal of Derivatives}\/},  {\it
  \bibinfo{volume}{3}\/}, \bibinfo{pages}{73--84}.
\bibitem[{LeCun et~al.(2015)LeCun, Bengio \& Hinton}]{lecun2015}
\bibinfo{author}{LeCun, Y.}, \bibinfo{author}{Bengio, Y.}, \&
  \bibinfo{author}{Hinton, G.} (\bibinfo{year}{2015}).
\newblock \bibinfo{title}{Deep learning}.
\newblock {\it \bibinfo{journal}{Nature}\/},  {\it \bibinfo{volume}{521}\/},
  \bibinfo{pages}{436--444}.
\bibitem[{{Meng} et~al.(2009){Meng}, {Dong} \& {Wong}}]{Meng2009}
\bibinfo{author}{{Meng}, K.}, \bibinfo{author}{{Dong}, Z.~Y.}, \&
  \bibinfo{author}{{Wong}, K.~P.} (\bibinfo{year}{2009}).
\newblock \bibinfo{title}{Self-adaptive radial basis function neural network
  for short-term electricity price forecasting}.
\newblock {\it \bibinfo{journal}{IET Generation, Transmission Distribution}\/},
   {\it \bibinfo{volume}{3}\/}, \bibinfo{pages}{325--335}.
\bibitem[{Nikolaev et~al.(2013)Nikolaev, Tino \& Smirnov}]{NTS2013}
\bibinfo{author}{Nikolaev, N.}, \bibinfo{author}{Tino, P.}, \&
  \bibinfo{author}{Smirnov, E.} (\bibinfo{year}{2013}).
\newblock \bibinfo{title}{Time-dependent series variance learning with
  recurrent mixture density networks}.
\newblock {\it \bibinfo{journal}{Neurocomputing}\/},  {\it
  \bibinfo{volume}{122}\/}, \bibinfo{pages}{501--512}.
\bibitem[{Nowotarski \& Weron(2018)}]{NowotarskiWeron2018}
\bibinfo{author}{Nowotarski, J.}, \& \bibinfo{author}{Weron, R.}
  (\bibinfo{year}{2018}).
\newblock \bibinfo{title}{Recent advances in electricity price forecasting: A
  review of probabilistic forecasting}.
\newblock {\it \bibinfo{journal}{Renewable and Sustainable Energy Reviews}\/},
  {\it \bibinfo{volume}{81}\/}, \bibinfo{pages}{1548--1568}.
\bibitem[{Ormoneit \& Neuneier(1996)}]{OrmoneitNeuneier1996}
\bibinfo{author}{Ormoneit, D.}, \& \bibinfo{author}{Neuneier, R.}
  (\bibinfo{year}{1996}).
\newblock \bibinfo{title}{Experiments in predicting the {G}erman stock index
  {DAX} with density estimating neural networks}.
\newblock In {\it \bibinfo{booktitle}{IEEE/IAFE 1996 Conference on
  Computational Intelligence for Financial Engineering (CIFEr)}\/} (pp.
  \bibinfo{pages}{66--71}).
\newblock \bibinfo{organization}{IEEE}.
\bibitem[{Patro \& Sahu(2015)}]{patro2015normalization}
\bibinfo{author}{Patro, S.}, \& \bibinfo{author}{Sahu, K.~K.}
  (\bibinfo{year}{2015}).
\newblock \bibinfo{title}{Normalization: A preprocessing stage}.
\newblock {\it \bibinfo{journal}{arXiv preprint arXiv:1503.06462}\/}, .
\bibitem[{Rasmussen \& Williams(2006)}]{Rasmussen06}
\bibinfo{author}{Rasmussen, C.}, \& \bibinfo{author}{Williams, C.}
  (\bibinfo{year}{2006}).
\newblock {\it \bibinfo{title}{Gaussian Processes for Machine Learning}\/}.
\newblock Adaptive Computation and Machine Learning.
\newblock \bibinfo{address}{Cambridge, MA, USA}: \bibinfo{publisher}{MIT
  Press}.
\bibitem[{Ripley(2007)}]{Ripley2007pattern}
\bibinfo{author}{Ripley, B.~D.} (\bibinfo{year}{2007}).
\newblock {\it \bibinfo{title}{Pattern recognition and neural networks}\/}.
\newblock \bibinfo{publisher}{Cambridge university press}.
\bibitem[{Rodrigues et~al.(2014)Rodrigues, Cardeira \& Calado}]{RCF2014}
\bibinfo{author}{Rodrigues, F.}, \bibinfo{author}{Cardeira, C.}, \&
  \bibinfo{author}{Calado, J. M.~F.} (\bibinfo{year}{2014}).
\newblock \bibinfo{title}{The daily and hourly energy consumption and load
  forecasting using artificial neural network method: a case study using a set
  of 93 households in portugal}.
\newblock {\it \bibinfo{journal}{Energy Procedia}\/},  {\it
  \bibinfo{volume}{62}\/}, \bibinfo{pages}{220--229}.
\bibitem[{{Shi} et~al.(2018){Shi}, {Xu} \& {Li}}]{Shi2018}
\bibinfo{author}{{Shi}, H.}, \bibinfo{author}{{Xu}, M.}, \&
  \bibinfo{author}{{Li}, R.} (\bibinfo{year}{2018}).
\newblock \bibinfo{title}{Deep learning for household load forecasting: A novel
  pooling deep {RNN}}.
\newblock {\it \bibinfo{journal}{IEEE Transactions on Smart Grid}\/},  {\it
  \bibinfo{volume}{9}\/}, \bibinfo{pages}{5271--5280}.
\bibitem[{Vossen et~al.(2018)Vossen, Feron \& Monti}]{VFM2018}
\bibinfo{author}{Vossen, J.}, \bibinfo{author}{Feron, B.}, \&
  \bibinfo{author}{Monti, A.} (\bibinfo{year}{2018}).
\newblock \bibinfo{title}{Probabilistic forecasting of household electrical
  load using artificial neural networks}.
\newblock In {\it \bibinfo{booktitle}{2018 IEEE International Conference on
  Probabilistic Methods Applied to Power Systems (PMAPS)}\/} (pp.
  \bibinfo{pages}{1--6}).

\end{thebibliography}
	
\end{document}